\documentclass[preprint,showpacs,superscriptaddress]{revtex4}
\usepackage[pdftex]{graphicx}
\usepackage{amssymb,amsfonts,amsmath,verbatim}
\usepackage{epstopdf}
\usepackage{setspace}

\begin{document}

\preprint{HEP/123-qed}
\title{Non-Markovian full counting statistics of cotunneling assisted
sequential tunneling in open quantum systems}
\author{Hai-Bin Xue}
\email{xuehaibin@tyut.edu.cn}
\affiliation{College of Physics and Optoelectronics, Taiyuan University of Technology,
Taiyuan, Shanxi 030024, China}
\author{Jiu-Qing Liang}
\affiliation{Institute of Theoretical Physics, Shanxi University, Taiyuan, Shanxi 030006,
China}
\author{Wu-Ming Liu}
\email{wmliu@iphy.ac.cn}
\affiliation{Beijing National Laboratory for Condensed Matter Physics, Institute of
Physics, Chinese Academy of Sciences, Beijing, 100190, China.}

\begin{abstract}
We develop a non-Markovian full counting statistics formalism taking into
account both the sequential tunneling and cotunneling based on the exact
particle number resolved time-convolutionless master equation and the
Rayleigh-Schr\"{o}dinger perturbation theory. Then, in the sequential
tunneling regime, we study the influences of the quantum coherence and the
cotunneling processes on the non-Markovian full counting statistics of
electron tunneling through an open quantum system, which consists of a
side-coupled double quantum-dot system weakly coupled to two electron
reservoirs. We demonstrate that, for the strong quantum-coherent
side-coupled double quantum-dot system, the competition or interplay between
the quantum coherence and the cotunneling processes, in the sequential
tunneling regime, determines whether the super-Poissonian distributions of
the shot noise, the skewness and the kurtosis take place (i.e., the Fano
factor is larger than one), and whether the sign transitions of the values
of the skewness and the kurtosis occur. These results suggest that, in the
sequential tunneling regime, it is necessary to consider the influences of
the quantum coherence and the cotunneling processes on the full counting
statistics in the open strong quantum-coherent quantum systems, which
provide a deeper insight into understanding of electron tunneling through
open quantum systems.
\end{abstract}

\date{\today}
\maketitle

\section{INTRODUCTION}

In an open quantum system, for an intermediate coupling strength between the
open quantum system and electron reservoir, the high-order tunneling
processes, i.e., the cotunneling processes can influence the electron
tunneling processes and bring out novel physical properties. Therefore, the
electron cotunneling in open quantum systems, especially quantum dot (QD)
systems, which is artificial molecules made from coupled QDs, and single
molecules have been extensively studied both experimentally \cite%
{FranceschiEx,LiuEx,SchleserEx,SigristEx,JoEx,RochEx} and theoretically \cite%
{HartmannTDia,GolovachTCAS,WeymannTDia,ElsteTDia,MisiornyTDia,BegemannTDia}.
Particularly, the shot noise \cite%
{OnacSEx,GustavssonSEx,ZhangSEx,ZarchinSEx,GustavssonSEx08,OkazakiSEx,SukhorukovSToff,ThielmannST,WeymannST,WeymannCAS,AghassiCAS,WeymannST10, WeymannST11,CarmiSTCAS,KaasbjergST,WrzesniewskiST}
and the full counting statistics (FCS) \cite%
{KaasbjergST,BraggioFCS,UtsumiFCS,EmaryFCS} of cotunneling in QD systems
have attracted considerable attention due to they can allow one to identify
the intrinsic properties of the QD systems and access the information of
electron correlation that cannot be obtained through the average current
measurements. For example, in the Coulomb blockade regime, in which the
transfer of electrons is dominated by cotunneling processes, it has been
demonstrated experimentally \cite%
{OnacSEx,GustavssonSEx,ZhangSEx,ZarchinSEx,GustavssonSEx08,OkazakiSEx} and
theoretically \cite%
{SukhorukovSToff,ThielmannST,WeymannST,WeymannCAS,AghassiCAS,WeymannST10,WeymannST11,CarmiSTCAS,KaasbjergST,WrzesniewskiST}
that the transport current displays super-Poissonian shot noise, which
indicates that the super-Poissonian distribution of transferred-electron
number and has a width being broader than its mean.

On the other hand, the quantum coherence, characterized by the off-diagonal
elements of reduced density matrix of the considered system, also plays an
important role in the electron tunneling through the strong quantum-coherent
systems \cite%
{SukhorukovSToff,Kieblichoff,Lindebaumoff,Urbanoff,Xueoff1,Xueoff2,Xueoff3,Xueoff4}%
. In particular, theoretical studies have demonstrated that the
non-Markovian effect of a strong quantum-coherent system plays an important
role in the non-equilibrium electron tunneling processes \cite{DMarcos11},
and manifests itself through the quantum coherence \cite{Xueoff3}.
Consequently, in the intermediate coupling strength case, the electron
tunneling through an open quantum system are mainly governed by the
competitions or interplays between the cotunneling, sequential tunneling and
quantum coherence. In the Coulomb blockade regime, it has been demonstrated
that the electron cotunneling processes play a crucial role; whereas that,
in the sequential tunneling regime where the transfer of electrons being
dominated by sequential tunneling, has a slightly influence on the
conduction and shot noise \cite%
{SchleserEx,WrzesniewskiST,DMarcos11,WeymannCAS0}. Outside the Coulomb
blockade regime, including the transition region from Coulomb blockade to
sequential tunneling and the sequential tunneling regime, theoretical
studies have demonstrated that the cotunneling assisted sequential tunneling
processes have an important influence on the conduction \cite%
{GolovachTCAS,AghassiCAS} and shot noise \cite{CarmiSTCAS,WeymannCAS}.
However, in the sequential tunneling regime, the influences of quantum
coherence and the cotunneling assisted sequential tunneling processes on the
non-Markovian FCS has not yet been revealed.

In this work, we derive a non-Markovian FCS formalism taking
into account both the sequential tunneling and cotunneling based on the
exact particle number resolved time-convolutionless (TCL) master equation
and the Rayleigh-Schr\"{o}dinger perturbation theory developed in references
\cite{Flindt01,Flindt02,Flindt03}. Then, in the sequential tunneling regime,
we study the influences of the quantum coherence and the cotunneling processes
on the non-Markovian FCS of electron tunneling through open quantum systems.
For the sake of discussion, the considered open quantum system consists of a
side-coupled double quantum-dot system weakly coupled to two electron
electrodes (reservoirs). Here, the corresponding quantum coherence can be tuned by
modulating the hopping strength between the two QDs relative to the coupling
of this QD molecule with the source and drain electrodes. It is numerically
demonstrated that, for the strong quantum-coherent side-coupled double QD
system, the competition or interplay between the cotunneling processes and
the quantum coherence, in the sequential tunneling regime, can take place in
a range of the QD-electrode coupling strength and has a remarkable
influence on the FCS. These characteristics depend on the temperature and the left-right
asymmetry of the QD-electrode coupling. Therefore, in the open strong
quantum-coherent quantum systems, it should be considered the effects of the
quantum coherence and the cotunneling processes on the FCS, even through the
electron tunneling is mainly dominated by the sequential tunneling processes.

\section{MODEL AND FORMALISM}

\subsection{Hamiltonian of open quantum system and TCL master equations}

We consider an open quantum system (OQS) weakly coupled to the two
electrodes (reservoirs), which is described by the following Hamiltonian
\begin{equation}
H=H_{\text{electrodes}}+H_{\text{OQS}}+H_{\text{hyb}}.  \label{Hamiltonian}
\end{equation}%
Here, the first term $H_{\text{electrodes}}=\sum_{\alpha ,k,\sigma
}\varepsilon _{\alpha k}a_{\alpha k\sigma }^{\dag }a_{\alpha k\sigma }$,
characterized by the two noninteracting reservoirs, stands for the
Hamiltonian of the two electrodes, with $\varepsilon_{\alpha k}$ being the
energy dispersion, and $a_{\alpha k\sigma }^{\dag }$ ($a_{\alpha k\sigma }$)
the creation (annihilation) operators in the $\alpha $ electrode. The second
term $H_{\text{OQS}}=H_{S}\left( d_{\mu }^{\dag },d_{\mu }\right) $, which
may contain vibrational or spin degrees of freedom and different types of
many-body interactions, represents the OQS Hamiltonian, where $d_{\mu
}^{\dag }$ ($d_{\mu }$) is the creation (annihilation) operator of electron
in a quantum state denoted by $\mu $. The third term $H_{\text{hyb}%
}=\sum_{\alpha ,\mu ,k}\left( t_{\alpha \mu k}d_{\mu }^{\dag }a_{\alpha \mu
k}+t_{\alpha \mu k}^{\ast }a_{\alpha \mu k}^{\dag }d_{\mu }\right) $, which
is assumed to be a sum of bilinear terms that each create an electron in the
OQS and annihilate one in the electrodes or vice versa, describes the
tunneling coupling between the OQS and the two electrodes.

Due to the OQS-electrode coupling being sufficiently weak, thus, $H_{\text{%
hyb}}$ can be treated perturbatively. In the interaction representation, the
equation of motion for the total density matrix reads%
\begin{equation}
\frac{\partial }{\partial t}\rho _{I}\left( t\right) =-i\left[ H_{\text{hyb}%
}^{I}\left( t\right) ,\rho ^{I}\left( t\right) \right] \equiv \mathcal{L}%
\left( t\right) \rho _{I}\left( t\right) ,  \label{interaction}
\end{equation}%
with%
\begin{equation*}
H_{\text{hyb}}^{I}\left( t\right) =\sum_{\alpha ,\mu }\left[ f_{\alpha \mu
}^{\dag }\left( t\right) d_{\mu }\left( t\right) +f_{\alpha \mu }\left(
t\right) d_{\mu }^{\dag }\left( t\right) \right]
\end{equation*}%
where%
\begin{equation}
f_{\alpha \mu }^{\dag }\left( t\right) =\sum_{k}t_{\alpha \mu k}^{\ast }\exp
\left( iH_{\text{electrodes}}t\right) a_{\alpha \mu k}^{\dag }\exp \left(
-iH_{\text{electrodes}}t\right)  \label{Interf}
\end{equation}
\begin{equation}
d_{\mu }\left( t\right) =\exp \left( iH_{\text{OQS}}t\right) d_{\mu }\exp
\left( -iH_{\text{OQS}}t\right)  \label{Interd}
\end{equation}%
To derive an exact equation of motion for the reduced density matrix $\rho
_{S}$ of the OQS, it is convenient to define a super-operator $\mathcal{P}$
according to%
\begin{equation}
\mathcal{P}\rho =\text{tr}_{B}\left[ \rho \right] \otimes \rho _{B}=\rho
_{S}\otimes \rho _{B},  \label{Pdefinition}
\end{equation}%
where $\rho _{B}$ is some fixed states of the two electrodes. Accordingly, a
complementary super-operator $\mathcal{Q},$%
\begin{equation}
\mathcal{Q}\rho =\rho -\mathcal{P}\rho .  \label{Qdefinition}
\end{equation}

For a factorizing initial condition $\rho \left( t_{0}\right) =\rho
_{S}\left( t_{0}\right) \otimes \rho _{B}$, $\mathcal{P}\rho \left(
t_{0}\right) =\rho \left( t_{0}\right) $, and then $\mathcal{Q}\rho \left(
t_{0}\right) =0$. Using the above TCL projection operator method, one can
obtain the second-order and the fourth-order TCL master equations \cite{book}%
\begin{equation}
\left. \frac{\partial }{\partial t}\mathcal{P}\rho \left( t\right)
\right\vert _{\text{sceond-order}}=\mathcal{K}_{2}\left( t\right) \mathcal{P}%
\rho \left( t\right) =\int_{-\infty }^{t}dt_{1}\mathcal{PL}\left( t\right)
\mathcal{L}\left( t_{1}\right) \mathcal{P}\rho \left( t\right)
\label{second-order}
\end{equation}%
\begin{eqnarray}
&&\left. \frac{\partial }{\partial t}\mathcal{P}\rho \left( t\right)
\right\vert _{\text{fourth-order}}  \notag \\
&=&\mathcal{K}_{4}\left( t\right) \mathcal{P}\rho \left( t\right)
=\int_{-\infty }^{t}dt_{1}\int_{-\infty }^{t_{1}}dt_{2}\int_{-\infty
}^{t_{2}}dt_{3}  \notag \\
&&\times \left[ \mathcal{PL}\left( t\right) \mathcal{L}\left( t_{1}\right)
\mathcal{L}\left( t_{2}\right) \mathcal{L}\left( t_{3}\right) \mathcal{P-PL}%
\left( t\right) \mathcal{L}\left( t_{1}\right) \mathcal{PL}\left(
t_{2}\right) \mathcal{L}\left( t_{3}\right) \mathcal{P}\right.  \notag \\
&&\left. -\mathcal{PL}\left( t\right) \mathcal{L}\left( t_{2}\right)
\mathcal{PL}\left( t_{1}\right) \mathcal{L}\left( t_{3}\right) \mathcal{P-PL}%
\left( t\right) \mathcal{L}\left( t_{3}\right) \mathcal{PL}\left(
t_{1}\right) \mathcal{L}\left( t_{2}\right) \mathcal{P}\right] \mathcal{P}%
\rho \left( t\right)  \label{four-order}
\end{eqnarray}%
Both Eqs. (\ref{second-order}) and (\ref{four-order}) are the starting point
of deriving the particle number resolved quantum master equation.

\subsection{The second-order particle number resolved TCL master equation}

In this subsection, we derive the second-order particle number resolved
quantum master equation based on Eq. (\ref{second-order}). Using both Eqs. (%
\ref{interaction}) and (\ref{Pdefinition}), Eq. (\ref{second-order}) can be
rewritten as
\begin{eqnarray}
&&\left. \frac{\partial }{\partial t}\rho _{I,S}\left( t\right) \right\vert
_{\text{sceond-order}}  \notag \\
&=&-\sum_{\alpha ij}\int_{-\infty }^{t}dt_{1}\text{tr}_{B}\left[ \rho
_{I,S}\left( t\right) \otimes \rho _{B}f_{\alpha j}^{\dag }\left(
t_{1}\right) d_{j}\left( t_{1}\right) d_{i}^{\dag }\left( t\right) f_{\alpha
i}\left( t\right) \right]  \notag \\
&&-\sum_{\alpha ij}\int_{-\infty }^{t}dt_{1}\text{tr}_{B}\left[ d_{i}^{\dag
}\left( t\right) f_{\alpha i}\left( t\right) f_{\alpha j}^{\dag }\left(
t_{1}\right) d_{j}\left( t_{1}\right) \rho _{I,S}\left( t\right) \otimes
\rho _{B}\right]  \notag \\
&&+\sum_{\alpha ij}\int_{-\infty }^{t}dt_{1}\text{tr}_{B}\left[ f_{\alpha
i}^{\dag }\left( t\right) d_{i}\left( t\right) \rho _{I,S}\left( t\right)
\otimes \rho _{B}d_{j}^{\dag }\left( t_{1}\right) f_{\alpha j}\left(
t_{1}\right) \right]  \notag \\
&&+\sum_{\alpha ij}\int_{-\infty }^{t}dt_{1}\text{tr}_{B}\left[ d_{i}^{\dag
}\left( t\right) f_{\alpha i}\left( t\right) \rho _{I,S}\left( t\right)
\otimes \rho _{B}f_{\alpha j}^{\dag }\left( t_{1}\right) d_{j}\left(
t_{1}\right) \right] +\text{H.c.}.  \label{InterEQM}
\end{eqnarray}

To fully describe the electron transport processes, the electron numbers,
which emitted from the source electrode and passed through the OQS and arrived at the drain electrode,
should be recorded. Following Refs. \cite{Lixq,WangSK}, the Hilbert subspace $B^{\left(n\right) }$ $\left(
n=1,2,...\right) $, which corresponds to $n$ electrons arriving at the
drain electrode and spanned by the product of all many-particle states
of the two electrodes, is introduced and formally denoted as $B^{\left( n\right)
}\equiv $ span$\left\{ \left\vert \Psi _{L}\right\rangle ^{\left( n\right)
}\otimes \left\vert \Psi _{R}\right\rangle ^{\left( n\right) }\right\} $.
Consequently, the entire Hilbert space of the two electrodes can be
expressed as $B=\oplus _{n}B^{\left( n\right) }$. With this classification
of the states of the two electrodes, the average over states in the entire
Hilbert space $B$ in Eq. (\ref{InterEQM}) should be replaced with the states
in the subspace $B^{\left( n\right) }$. Then, Eq. (\ref{InterEQM}) can be
expressed as a conditional TCL master equation
\begin{eqnarray}
&&\left. \frac{\partial }{\partial t}\rho _{I,S}^{\left( n\right) }\left(
t\right) \right\vert _{\text{sceond-order}}  \notag \\
&=&-\sum_{\alpha ij}\int_{-\infty }^{t}dt_{1}\text{tr}_{B^{\left( n\right) }}%
\left[ \rho _{I,S}\left( t\right) \otimes \rho _{B}f_{\alpha j}^{\dag
}\left( t_{1}\right) d_{j}\left( t_{1}\right) d_{i}^{\dag }\left( t\right)
f_{\alpha i}\left( t\right) \right]  \notag \\
&&-\sum_{\alpha ij}\int_{-\infty }^{t}dt_{1}\text{tr}_{B^{\left( n\right) }}%
\left[ d_{i}^{\dag }\left( t\right) f_{\alpha i}\left( t\right) f_{\alpha
j}^{\dag }\left( t_{1}\right) d_{j}\left( t_{1}\right) \rho _{I,S}\left(
t\right) \otimes \rho _{B}\right]  \notag \\
&&+\sum_{\alpha ij}\int_{-\infty }^{t}dt_{1}\text{tr}_{B^{\left( n\right) }}%
\left[ f_{\alpha j}^{\dag }\left( t_{1}\right) d_{j}\left( t_{1}\right) \rho
_{I,S}\left( t\right) \otimes \rho _{B}d_{i}^{\dag }\left( t\right)
f_{\alpha i}\left( t\right) \right]  \notag \\
&&+\sum_{\alpha ij}\int_{-\infty }^{t}dt_{1}\text{tr}_{B^{\left( n\right) }}%
\left[ d_{i}^{\dag }\left( t\right) f_{\alpha i}\left( t\right) \rho
_{I,S}\left( t\right) \otimes \rho _{B}f_{\alpha j}^{\dag }\left(
t_{1}\right) d_{j}\left( t_{1}\right) \right] +\text{H.c.}.  \label{ndensity}
\end{eqnarray}

Before proceeding, two physical considerations are implemented. (i) Instead
of the conventional Born approximation for the entire density matrix $\rho
_{T}\left( t\right) \simeq \rho \left( t\right) \otimes \rho _{B}$, the
ansatz $\rho ^{I}\left( t\right) \simeq \rho ^{\left( n\right) }\left(
t\right) \otimes \rho _{B}^{\left( n\right) }$ is proposed, where $\rho
_{B}^{\left( n\right) }$ being the density operator of two electrodes
associated with $n$ electrons arriving at the drain electrode. With this
ansatz for the entire density operator (i.e., tracing over the subspace $%
B^{\left( n\right) }$), Eq. (\ref{ndensity}) can be reexpressed as
\begin{eqnarray}
&&\left. \frac{\partial }{\partial t}\rho _{I,S}^{\left( n\right) }\left(
t\right) \right\vert _{\text{sceond-order}}  \notag \\
&=&-\sum_{\alpha ij}\int_{-\infty }^{t}dt_{1}\text{tr}_{B^{\left( n\right) }}%
\left[ f_{\alpha j}^{\dag }\left( t_{1}\right) f_{\alpha i}\left( t\right)
\rho _{B}\right] \rho _{I,S}^{\left( n\right) }\left( t\right) d_{j}\left(
t_{1}\right) d_{i}^{\dag }\left( t\right)  \notag \\
&&-\sum_{\alpha ij}\int_{-\infty }^{t}dt_{1}\text{tr}_{B^{\left( n\right) }}%
\left[ f_{\alpha i}\left( t\right) f_{\alpha j}^{\dag }\left( t_{1}\right)
\rho _{B}\right] d_{i}^{\dag }\left( t\right) d_{j}\left( t_{1}\right) \rho
_{I,S}^{\left( n\right) }\left( t\right)  \notag \\
&&+\sum_{ij}\int_{-\infty }^{t}dt_{1}\text{tr}_{B^{\left( n\right) }}\left[
f_{Li}\left( t\right) f_{Lj}^{\dag }\left( t_{1}\right) \rho _{B}\right]
d_{j}\left( t_{1}\right) \rho _{I,S}^{\left( n\right) }\left( t\right)
d_{i}^{\dag }\left( t\right)  \notag \\
&&+\sum_{ij}\int_{-\infty }^{t}dt_{1}\text{tr}_{B^{\left( n\right) }}\left[
f_{Ri}\left( t\right) f_{Rj}^{\dag }\left( t_{1}\right) \rho _{B}\right]
d_{j}\left( t_{1}\right) \rho _{I,S}^{\left( n-1\right) }\left( t\right)
d_{i}^{\dag }\left( t\right)  \notag \\
&&+\sum_{ij}\int_{-\infty }^{t}dt_{1}\text{tr}_{B^{\left( n\right) }}\left[
f_{Lj}^{\dag }\left( t_{1}\right) f_{Li}\left( t\right) \rho _{B}\right]
d_{i}^{\dag }\left( t\right) \rho _{I,S}^{\left( n\right) }\left( t\right)
d_{j}\left( t_{1}\right)  \notag \\
&&+\sum_{ij}\int_{-\infty }^{t}dt_{1}\text{tr}_{B^{\left( n\right) }}\left[
f_{Rj}^{\dag }\left( t_{1}\right) f_{Ri}\left( t\right) \rho _{B}\right]
d_{i}^{\dag }\left( t\right) \rho _{I,S}^{\left( n+1\right) }\left( t\right)
d_{j}\left( t_{1}\right) +\text{H.c.}.  \label{ndensitymodified}
\end{eqnarray}%
Here, we have used the orthogonality between the states in different
subspaces. (ii) The extra electrons arriving at the drain electrode will
flow back into the source electrode via the external closed transport
circuit. Additionally, the rapid relaxation processes in the electrodes will
bring the electrodes to the local thermal equilibrium states quickly, which
are determined by the chemical potentials. After the procedure done in Eq. (%
\ref{ndensitymodified}), the density matrices of two electrodes $\rho
_{B}^{\left( n\right) }$ and $\rho _{B}^{\left( n\pm 1\right) }$ should be
replaced by $\rho _{B}^{\left( 0\right) }$. In the Schr\"{o}dinger
representation, Eq. (\ref{ndensitymodified}) can be written as
\begin{eqnarray}
&&\left. \frac{\partial }{\partial t}\rho _{S}^{\left( n\right) }\left(
t\right) \right\vert _{\text{sceond-order}}  \notag \\
&=&-i\left[ H_{S},\rho _{S}^{\left( n\right) }\left( t\right) \right]  \notag
\\
&&-\sum_{\alpha ij}\int_{-\infty }^{t}dt_{1}C_{\alpha ji}^{\left( +\right)
}\left( t_{1}-t\right) \rho _{S}^{\left( n\right) }\left( t\right)
e^{-iH_{S}\left( t-t_{1}\right) }d_{j}e^{iH_{S}\left( t-t_{1}\right)
}d_{i}^{\dag }  \notag \\
&&-\sum_{\alpha ij}\int_{-\infty }^{t}dt_{1}C_{\alpha ij}^{\left( -\right)
}\left( t-t_{1}\right) d_{i}^{\dag }e^{-iH_{S}\left( t-t_{1}\right)
}d_{j}e^{iH_{S}\left( t-t_{1}\right) }\rho _{S}^{\left( n\right) }\left(
t\right)  \notag \\
&&+\sum_{ij}\int_{-\infty }^{t}dt_{1}C_{Lij}^{\left( -\right) }\left(
t-t_{1}\right) e^{-iH_{S}\left( t-t_{1}\right) }d_{j}e^{iH_{S}\left(
t-t_{1}\right) }\rho _{S}^{\left( n\right) }\left( t\right) d_{i}^{\dag }
\notag \\
&&+\sum_{ij}\int_{-\infty }^{t}dt_{1}C_{Rij}^{\left( -\right) }\left(
t-t_{1}\right) e^{-iH_{S}\left( t-t_{1}\right) }d_{j}e^{iH_{S}\left(
t-t_{1}\right) }\rho _{S}^{\left( n-1\right) }\left( t\right) d_{i}^{\dag }
\notag \\
&&+\sum_{ij}\int_{-\infty }^{t}dt_{1}C_{Lji}^{\left( +\right) }\left(
t_{1}-t\right) d_{i}^{\dag }\rho _{S}^{\left( n\right) }\left( t\right)
e^{-iH_{S}\left( t-t_{1}\right) }d_{j}e^{iH_{S}\left( t-t_{1}\right) }
\notag \\
&&+\sum_{ij}\int_{-\infty }^{t}dt_{1}C_{Rji}^{\left( +\right) }\left(
t_{1}-t\right) d_{i}^{\dag }\rho _{S}^{\left( n+1\right) }\left( t\right)
e^{-iH_{S}\left( t-t_{1}\right) }d_{j}e^{iH_{S}\left( t-t_{1}\right) }+\text{%
H.c.}.  \label{nQME}
\end{eqnarray}%
where the correlation functions are defined as
\begin{equation}
C_{\alpha ij}^{\left( +\right) }\left( t-t_{1}\right) =\text{tr}_{R}\left[
f_{\alpha i}^{\dag }\left( t\right) f_{\alpha j}\left( t_{1}\right) \rho _{B}%
\right] =\left\langle f_{\alpha i}^{\dagger }\left( t\right) f_{\alpha
j}\left( t_{1}\right) \right\rangle ,  \label{Correfun01}
\end{equation}%
\begin{equation}
C_{\alpha ij}^{\left( -\right) }\left( t-t_{1}\right) =\text{tr}_{R}\left[
f_{\alpha i}\left( t\right) f_{\alpha j}^{\dag }\left( t_{1}\right) \rho _{B}%
\right] =\left\langle f_{\alpha i}\left( t\right) f_{\alpha j}^{\dagger
}\left( t_{1}\right) \right\rangle .  \label{Correfun02}
\end{equation}%
Introducing the following super-operators
\begin{equation}
A_{\alpha i}^{\left( +\right) }\left( t\right) =\sum_{j}\int_{-\infty
}^{t}dt_{1}C_{\alpha ji}^{\left( +\right) }\left( t_{1}-t\right)
e^{-iH_{S}\left( t-t_{1}\right) }d_{j}e^{iH_{S}\left( t-t_{1}\right) },
\label{superoperator01}
\end{equation}%
\begin{equation}
A_{\alpha i}^{\left( -\right) }\left( t\right) =\sum_{j}\int_{-\infty
}^{t}dt_{1}C_{\alpha ij}^{\left( -\right) }\left( t-t_{1}\right)
e^{-iH_{S}\left( t-t_{1}\right) }d_{j}e^{iH_{S}\left( t-t_{1}\right) },
\label{superoperator02}
\end{equation}%
then, Eq. (\ref{nQME}) can be rewritten as a compact form
\begin{eqnarray}
&&\left. \frac{\partial }{\partial t}\rho _{S}^{\left( n\right) }\left(
t\right) \right\vert _{\text{sceond-order}}  \notag \\
&=&-i\left[ H_{S},\rho _{S}^{\left( n\right) }\left( t\right) \right]  \notag
\\
&&-\sum_{i}\left\{ \rho _{S}^{\left( n\right) }\left( t\right) A_{i}^{\left(
+\right) }\left( t\right) d_{i}^{\dag }+d_{i}^{\dag }A_{i}^{\left( -\right)
}\left( t\right) \rho _{S}^{\left( n\right) }\left( t\right) -A_{Li}^{\left(
-\right) }\left( t\right) \rho _{S}^{\left( n\right) }\left( t\right)
d_{i}^{\dag }\right.  \notag \\
&&\left. -A_{Ri}^{\left( -\right) }\left( t\right) \rho _{S}^{\left(
n-1\right) }\left( t\right) d_{i}^{\dag }-d_{i}^{\dag }\rho _{S}^{\left(
n\right) }\left( t\right) A_{Li}^{\left( +\right) }\left( t\right)
-d_{i}^{\dag }\rho _{S}^{\left( n+1\right) }\left( t\right) A_{Ri}^{\left(
+\right) }\left( t\right) +\text{H.c.}\right\} .  \label{nQMEfinal}
\end{eqnarray}%
where $A_{i}^{\left( \pm \right) }\left( t\right) =\sum_{\alpha }A_{\alpha
i}^{\left( \pm \right) }\left( t\right) $. The equation (\ref{nQMEfinal}) is
the starting point of the non-Markovian FCS calculation taking into account
the sequential tunneling processes only.

\subsection{The fourth-order particle number resolved TCL master equation}

In this subsection, we derive the fourth-order particle number resolved
quantum master equation based on Eq. (\ref{four-order}). Using both Eqs. (%
\ref{interaction}) and (\ref{Pdefinition}), Eq. (\ref{four-order}) can be
expressed as%
\begin{eqnarray}
&&\left. \frac{\partial }{\partial t}\rho _{I,S}\left( t\right) \right\vert
_{\text{fourth-order}}=\int_{-\infty }^{t}dt_{1}\int_{-\infty
}^{t_{1}}dt_{2}\int_{-\infty }^{t_{2}}dt_{3}\sum\limits_{ijkl}  \notag \\
&&\times \left\{ \text{tr}_{B}\left[ H_{I}\left( t\right) ,\left[
H_{I}\left( t_{1}\right) ,\left[ H_{I}\left( t_{2}\right) ,\left[
H_{I}\left( t_{3}\right) ,\rho _{S}\otimes \rho _{B}\right] \right] \right] %
\right] \right.  \notag \\
&&-\text{tr}_{B}\left[ H_{I}\left( t\right) ,\left[ H_{I}\left( t_{1}\right)
,\text{tr}_{B}\left[ H_{I}\left( t_{2}\right) ,\left[ H_{I}\left(
t_{3}\right) ,\rho _{S}\otimes \rho _{B}\right] \right] \otimes \rho _{B}%
\right] \right]  \notag \\
&&-\text{tr}_{B}\left[ H_{I}\left( t\right) ,\left[ H_{I}\left( t_{2}\right)
,\text{tr}_{B}\left[ H_{I}\left( t_{1}\right) ,\left[ H_{I}\left(
t_{3}\right) ,\rho _{S}\otimes \rho _{B}\right] \right] \otimes \rho _{B}%
\right] \right]  \notag \\
&&-\text{tr}_{B}\left[ H_{I}\left( t\right) ,\left[ H_{I}\left( t_{3}\right)
,\text{tr}_{B}\left[ H_{I}\left( t_{1}\right) ,\left[ H_{I}\left(
t_{2}\right) ,\rho _{S}\otimes \rho _{B}\right] \right] \otimes \rho _{B}%
\right] \right]  \label{four-01}
\end{eqnarray}

To further facilitate this derivation, the tunneling coupling between the
OQS and the two electrodes $H_{\text{hyb}}^{I}\left( t\right) $ is
rewritten as the following equation%
\begin{equation}
H_{\text{hyb}}^{I}\left( t\right) =\sum_{\alpha ,\mu }F_{\alpha \mu }\left(
t\right) D_{\mu }\left( t\right) ,  \label{hypmpdified}
\end{equation}%
with%
\begin{equation*}
F_{\alpha \mu }\left( t\right) =\sum_{k}e^{iH_{\text{electrodes}}t}F_{\alpha
\mu k}e^{-iH_{\text{electrodes}}t},
\end{equation*}%
\begin{equation*}
D_{\mu }\left( t\right) =e^{iH_{\text{dot}}t}D_{\mu }e^{-iH_{\text{dot}}t},
\end{equation*}%
where $F_{\alpha \mu k}=t_{\alpha \mu k}a_{\alpha \mu k}+t_{\alpha \mu
k}^{\ast }a_{\alpha \mu k}^{\dag }$, $D_{\mu }=d_{\mu }+d_{\mu }^{\dag }$.
Inserting Eq. (\ref{hypmpdified}) into Eq. (\ref{four-01}), one can obtain
\cite{book}%
\begin{eqnarray}
&&\left. \frac{\partial }{\partial t}\rho _{I,S}\left( t\right) \right\vert
_{\text{fourth-order}}  \notag \\
&=&\int_{-\infty }^{t}dt_{1}\int_{-\infty }^{t_{1}}dt_{2}\int_{-\infty
}^{t_{2}}dt_{3}\sum\limits_{ijkl}\left\{ C_{02}C_{13}\left[ \widehat{0},%
\left[ \widehat{1},\widehat{2}\right] \widehat{3}\rho _{S}\right] \right.
\notag \\
&&-C_{02}C_{31}\left[ \widehat{0},\left[ \widehat{1},\widehat{2}\right] \rho
_{S}\widehat{3}\right] +C_{03}C_{12}\left[ \widehat{0},\left[ \widehat{1}%
\widehat{2},\widehat{3}\right] \rho _{S}\right]  \notag \\
&&\left. -C_{03}C_{12}\left[ \widehat{0},\left[ \widehat{2},\widehat{3}%
\right] \rho _{S}\widehat{1}\right] -C_{03}C_{21}\left[ \widehat{0},\left[
\widehat{1},\widehat{3}\right] \rho _{S}\widehat{2}\right] \right\} +\text{%
H.c.},  \label{four-02}
\end{eqnarray}%
with%
\begin{equation*}
C_{02}=\sum_{\alpha ik}\text{tr}_{B}\left[ F_{\alpha i}\left( t\right)
F_{\alpha k}\left( t_{2}\right) \right] \text{, }C_{03}=\sum_{\alpha il}%
\text{tr}_{B}\left[ F_{\alpha i}\left( t\right) F_{\alpha l}\left(
t_{3}\right) \right] \text{,}
\end{equation*}%
\begin{equation*}
C_{12}=\sum_{\alpha jk}\text{tr}_{B}\left[ F_{\alpha j}\left( t_{1}\right)
F_{\alpha k}\left( t_{2}\right) \right] \text{, }C_{21}=\sum_{\alpha kj}%
\text{tr}_{B}\left[ F_{\alpha k}\left( t_{2}\right) F_{\alpha j}\left(
t_{1}\right) \right] \text{,}
\end{equation*}%
\begin{equation*}
C_{13}=\sum_{\alpha jl}\text{tr}_{B}\left[ F_{\alpha j}\left( t_{1}\right)
F_{\alpha l}\left( t_{3}\right) \right] \text{, }C_{31}=\sum_{\alpha lj}%
\text{tr}_{B}\left[ F_{\alpha l}\left( t_{3}\right) F_{\alpha j}\left(
t_{1}\right) \right] \text{,}
\end{equation*}%
\begin{equation*}
\widehat{0}=D_{i}\left( t\right) \text{, }\widehat{1}=D_{j}\left(
t_{1}\right) \text{, }\widehat{2}=D_{k}\left( t_{2}\right) \text{, }\widehat{%
3}=D_{l}\left( t_{3}\right) .
\end{equation*}

In the Schr\"{o}dinger representation, Eq. (\ref{four-02}) can be expressed
as
\begin{eqnarray}
&&\left. \frac{\partial \rho _{S}\left( t\right) }{\partial t}\right\vert _{%
\text{fourth-order}}  \notag \\
&&=-i\left[ H_{S},\rho _{S}\left( t\right) \right] +\int_{-\infty
}^{t}dt_{1}\int_{-\infty }^{t_{1}}dt_{2}\int_{-\infty
}^{t_{2}}dt_{3}\sum_{ijkl}\left[ I_{1}+I_{2}+II_{1}+II_{2}+II_{3}+\text{H.c.}%
\right] ,  \label{four}
\end{eqnarray}%
with%
\begin{eqnarray}
&&I_{1}=C_{02}C_{13}D_{i}e^{-i\mathcal{L}_{S}\left( t-t_{1}\right)
}D_{j}e^{-i\mathcal{L}_{S}\left( t-t_{2}\right) }D_{k}e^{-i\mathcal{L}%
_{S}\left( t-t_{3}\right) }D_{l}\rho _{S}\left( t\right)  \notag \\
&&+C_{02}C_{13}e^{-i\mathcal{L}_{S}\left( t-t_{2}\right) }D_{k}e^{-i\mathcal{%
L}_{S}\left( t-t_{1}\right) }D_{j}e^{-i\mathcal{L}_{S}\left( t-t_{3}\right)
}D_{l}\rho _{S}\left( t\right) D_{i}  \notag \\
&&-C_{02}C_{13}D_{i}e^{-i\mathcal{L}_{S}\left( t-t_{2}\right) }D_{k}e^{-i%
\mathcal{L}_{S}\left( t-t_{1}\right) }D_{j}e^{-i\mathcal{L}_{S}\left(
t-t_{3}\right) }D_{l}\rho _{S}\left( t\right)  \notag \\
&&-C_{02}C_{13}e^{-i\mathcal{L}_{S}\left( t-t_{1}\right) }D_{j}e^{-i\mathcal{%
L}_{S}\left( t-t_{2}\right) }D_{k}e^{-i\mathcal{L}_{S}\left( t-t_{3}\right)
}D_{l}\rho _{S}\left( t\right) D_{i},  \label{I1}
\end{eqnarray}%
\begin{eqnarray}
&&I_{2}=C_{02}C_{31}e^{-i\mathcal{L}_{S}\left( t-t_{1}\right) }D_{j}e^{-i%
\mathcal{L}_{S}\left( t-t_{2}\right) }D_{k}\rho _{S}\left( t\right) e^{-i%
\mathcal{L}_{S}\left( t-t_{3}\right) }D_{l}D_{i}  \notag \\
&&+C_{02}C_{31}D_{i}e^{-i\mathcal{L}_{S}\left( t-t_{2}\right) }D_{k}e^{-i%
\mathcal{L}_{S}\left( t-t_{1}\right) }D_{j}\rho _{S}\left( t\right) e^{-i%
\mathcal{L}_{S}\left( t-t_{3}\right) }D_{l}  \notag \\
&&-C_{02}C_{31}e^{-i\mathcal{L}_{S}\left( t-t_{1}\right) }D_{j}e^{-i\mathcal{%
L}_{S}\left( t-t_{2}\right) }D_{k}\rho _{S}\left( t\right) e^{-i\mathcal{L}%
_{S}\left( t-t_{3}\right) }D_{l}  \notag \\
&&-C_{02}C_{31}e^{-i\mathcal{L}_{S}\left( t-t_{2}\right) }D_{k}e^{-i\mathcal{%
L}_{S}\left( t-t_{1}\right) }D_{j}\rho _{S}\left( t\right) e^{-i\mathcal{L}%
_{S}\left( t-t_{3}\right) }D_{l}D_{i},  \label{I2}
\end{eqnarray}%
\begin{eqnarray}
&&II_{1}=C_{03}C_{12}D_{i}e^{-i\mathcal{L}_{S}\left( t-t_{1}\right)
}D_{j}e^{-i\mathcal{L}_{S}\left( t-t_{2}\right) }D_{k}e^{-i\mathcal{L}%
_{S}\left( t-t_{3}\right) }D_{l}\rho _{S}\left( t\right)  \notag \\
&&+C_{03}C_{12}e^{-i\mathcal{L}_{S}\left( t-t_{3}\right) }D_{l}e^{-i\mathcal{%
L}_{S}\left( t-t_{1}\right) }D_{j}e^{-i\mathcal{L}_{S}\left( t-t_{2}\right)
}D_{k}\rho _{S}\left( t\right) D_{i}  \notag \\
&&-C_{03}C_{12}D_{i}e^{-i\mathcal{L}_{S}\left( t-t_{3}\right) }D_{l}e^{-i%
\mathcal{L}_{S}\left( t-t_{1}\right) }D_{j}e^{-i\mathcal{L}_{S}\left(
t-t_{2}\right) }D_{k}\rho _{S}\left( t\right)  \notag \\
&&-C_{03}C_{12}e^{-i\mathcal{L}_{S}\left( t-t_{1}\right) }D_{j}e^{-i\mathcal{%
L}_{S}\left( t-t_{2}\right) }D_{k}e^{-i\mathcal{L}_{S}\left( t-t_{3}\right)
}D_{l}\rho _{S}\left( t\right) D_{i},  \label{II1}
\end{eqnarray}%
\begin{eqnarray}
&&II_{2}=C_{03}C_{12}D_{i}e^{-i\mathcal{L}_{S}\left( t-t_{3}\right)
}D_{l}e^{-i\mathcal{L}_{S}\left( t-t_{2}\right) }D_{k}\rho _{S}\left(
t\right) e^{-i\mathcal{L}_{S}\left( t-t_{1}\right) }D_{j}  \notag \\
&&+C_{03}C_{12}e^{-i\mathcal{L}_{S}\left( t-t_{2}\right) }D_{k}e^{-i\mathcal{%
L}_{S}\left( t-t_{3}\right) }D_{l}\rho _{S}\left( t\right) e^{-i\mathcal{L}%
_{S}\left( t-t_{1}\right) }D_{j}D_{i}  \notag \\
&&-C_{03}C_{12}D_{i}e^{-i\mathcal{L}_{S}\left( t-t_{2}\right) }D_{k}e^{-i%
\mathcal{L}_{S}\left( t-t_{3}\right) }D_{l}\rho _{S}\left( t\right) e^{-i%
\mathcal{L}_{S}\left( t-t_{1}\right) }D_{j}  \notag \\
&&-C_{03}C_{12}e^{-i\mathcal{L}_{S}\left( t-t_{3}\right) }D_{l}e^{-i\mathcal{%
L}_{S}\left( t-t_{2}\right) }D_{k}\rho _{S}\left( t\right) e^{-i\mathcal{L}%
_{S}\left( t-t_{1}\right) }D_{j}D_{i},  \label{II2}
\end{eqnarray}%
\begin{eqnarray}
&&II_{3}=C_{03}C_{21}D_{i}e^{-i\mathcal{L}_{S}\left( t-t_{3}\right)
}D_{l}e^{-i\mathcal{L}_{S}\left( t-t_{1}\right) }D_{j}\rho _{S}\left(
t\right) e^{-i\mathcal{L}_{S}\left( t-t_{2}\right) }D_{k}  \notag \\
&&+C_{03}C_{21}e^{-i\mathcal{L}_{S}\left( t-t_{1}\right) }D_{j}e^{-i\mathcal{%
L}_{S}\left( t-t_{3}\right) }D_{l}\rho _{S}\left( t\right) e^{-i\mathcal{L}%
_{S}\left( t-t_{2}\right) }D_{k}D_{i}  \notag \\
&&-C_{03}C_{21}D_{i}e^{-i\mathcal{L}_{S}\left( t-t_{1}\right) }D_{j}e^{-i%
\mathcal{L}_{S}\left( t-t_{3}\right) }D_{l}\rho _{S}\left( t\right) e^{-i%
\mathcal{L}_{S}\left( t-t_{2}\right) }D_{k}  \notag \\
&&-C_{03}C_{21}e^{-i\mathcal{L}_{S}\left( t-t_{3}\right) }D_{l}e^{-i\mathcal{%
L}_{S}\left( t-t_{1}\right) }D_{j}\rho _{S}\left( t\right) e^{-i\mathcal{L}%
_{S}\left( t-t_{2}\right) }D_{k}D_{i},  \label{II3}
\end{eqnarray}%
Here, we define the super-operator $\mathcal{L}_{S}$ as $%
e^{-iH_{S}t}Oe^{iH_{S}t}\equiv e^{-i\mathcal{L}_{S}t}O$. Now, we derive the
fourth-order particle number resolved quantum master equation based on Eq. (%
\ref{four}). Without loss of generality, we consider the case of Eq. (\ref%
{I1}). Considering the Hamiltonian $H_{\text{hyb}%
}^{I}\left( t\right) $ $=\sum_{\alpha ,\mu }\left[ f_{\alpha \mu }^{\dag
}\left( t\right) d_{\mu }\left( t\right) +f_{\alpha \mu }\left( t\right)
d_{\mu }^{\dag }\left( t\right) \right] $, the $C_{02},C_{13}$ and $%
D_{i,j,k,l}$ have the following forms%
\begin{equation}
C_{02}^{\left( +\right) }=\sum_{\alpha ik}\text{tr}_{B}\left[ f_{\alpha
i}^{\dag }\left( t\right) f_{\alpha k}\left( t_{2}\right) \right]
,D_{i}=d_{i},D_{k}=d_{k}^{\dag },  \label{c02add}
\end{equation}%
\begin{equation}
C_{02}^{\left( -\right) }=\sum_{\alpha ik}\text{tr}_{B}\left[ f_{\alpha
i}\left( t\right) f_{\alpha k}^{\dag }\left( t_{2}\right) \right]
,D_{i}=d_{i}^{\dag },D_{k}=d_{k},  \label{c02sub}
\end{equation}%
\begin{equation}
C_{13}^{\left( +\right) }=\sum_{\alpha jl}\text{tr}_{B}\left[ f_{\alpha
j}^{\dag }\left( t_{1}\right) f_{\alpha l}\left( t_{3}\right) \right]
,D_{j}=d_{j},D_{l}=d_{l}^{\dag },  \label{c13add}
\end{equation}%
\begin{equation}
C_{13}^{\left( -\right) }=\sum_{\alpha jl}\text{tr}_{B}\left[ f_{\alpha
j}\left( t_{1}\right) f_{\alpha l}^{\dag }\left( t_{3}\right) \right]
,D_{j}=d_{j}^{\dag },D_{l}=d_{l},  \label{c13sub}
\end{equation}%
respectively. Therefore, the particle number resolved formation of Eq. (\ref%
{I1}) can be expressed as follows
\begin{equation}
I_{1}=I_{1,n-1}+I_{1,n}+I_{1,n+1},  \label{Imodified}
\end{equation}%
with%
\begin{eqnarray}
&&I_{1,n-1}=  \notag \\
&&+C_{R,0,2}^{\left( -\right) }C_{L,1,3}^{\left( -\right) }A_{k}A_{j}^{\dag
}A_{l}\rho _{S}^{\left( n-1\right) }\left( t\right) d_{i}^{\dagger
}+C_{R,0,2}^{\left( -\right) }C_{R,1,3}^{\left( -\right) }A_{k}A_{j}^{\dag
}A_{l}\rho _{S}^{\left( n-1\right) }\left( t\right) d_{i}^{\dagger }  \notag
\\
&&-C_{R,0,2}^{\left( -\right) }C_{L,1,3}^{\left( -\right) }A_{j}^{\dag
}A_{k}A_{l}\rho _{S}^{\left( n-1\right) }\left( t\right) d_{i}^{\dagger
}-C_{R,0,2}^{\left( -\right) }C_{R,1,3}^{\left( -\right) }A_{j}^{\dag
}A_{k}A_{l}\rho _{S}^{\left( n-1\right) }\left( t\right) d_{i}^{\dagger }
\notag \\
&&+C_{R,0,2}^{\left( -\right) }C_{L,1,3}^{\left( +\right)
}A_{k}A_{j}A_{l}^{\dag }\rho _{S}^{\left( n-1\right) }\left( t\right)
d_{i}^{\dagger }+C_{R,0,2}^{\left( -\right) }C_{R,1,3}^{\left( +\right)
}A_{k}A_{j}A_{l}^{\dag }\rho _{S}^{\left( n-1\right) }\left( t\right)
d_{i}^{\dagger }  \notag \\
&&-C_{R,0,2}^{\left( -\right) }C_{L,1,3}^{\left( +\right)
}A_{j}A_{k}A_{l}^{\dag }\rho _{S}^{\left( n-1\right) }\left( t\right)
d_{i}^{\dagger }-C_{R,0,2}^{\left( -\right) }C_{R,1,3}^{\left( +\right)
}A_{j}A_{k}A_{l}^{\dag }\rho _{S}^{\left( n-1\right) }\left( t\right)
d_{i}^{\dagger }  \label{I02}
\end{eqnarray}%
\begin{eqnarray}
&&I_{1,n}=  \notag \\
&&+C_{0,2}^{\left( -\right) }C_{1,3}^{\left( -\right) }d_{i}^{\dagger
}A_{j}^{\dag }A_{k}A_{l}\rho _{S}^{\left( n\right) }\left( t\right)
+C_{0,2}^{\left( -\right) }C_{1,3}^{\left( +\right) }d_{i}^{\dagger
}A_{j}A_{k}A_{l}^{\dag }\rho _{S}^{\left( n\right) }\left( t\right)  \notag
\\
&&-C_{0,2}^{\left( -\right) }C_{1,3}^{\left( -\right) }d_{i}^{\dagger
}A_{k}A_{j}^{\dag }A_{l}\rho _{S}^{\left( n\right) }\left( t\right)
-C_{0,2}^{\left( -\right) }C_{1,3}^{\left( +\right) }d_{i}^{\dagger
}A_{k}A_{j}A_{l}^{\dag }\rho _{S}^{\left( n\right) }\left( t\right)  \notag
\\
&&+C_{0,2}^{\left( +\right) }C_{1,3}^{\left( -\right) }d_{i}A_{j}^{\dag
}A_{k}^{\dag }A_{l}\rho _{S}^{\left( n\right) }\left( t\right)
+C_{0,2}^{\left( +\right) }C_{1,3}^{\left( +\right) }d_{i}A_{j}A_{k}^{\dag
}A_{l}^{\dag }\rho _{S}^{\left( n\right) }\left( t\right)  \notag \\
&&-C_{0,2}^{\left( +\right) }C_{1,3}^{\left( -\right) }d_{i}A_{k}^{\dag
}A_{j}^{\dag }A_{l}\rho _{S}^{\left( n\right) }\left( t\right)
-C_{0,2}^{\left( +\right) }C_{1,3}^{\left( +\right) }d_{i}A_{k}^{\dag
}A_{j}A_{l}^{\dag }\rho _{S}^{\left( n\right) }\left( t\right)  \notag \\
&&+C_{L,0,2}^{\left( -\right) }C_{L,1,3}^{\left( -\right) }A_{k}A_{j}^{\dag
}A_{l}\rho _{S}^{\left( n\right) }\left( t\right) d_{i}^{\dagger
}+C_{L,0,2}^{\left( -\right) }C_{R,1,3}^{\left( -\right) }A_{k}A_{j}^{\dag
}A_{l}\rho _{S}^{\left( n\right) }\left( t\right) d_{i}^{\dagger }  \notag \\
&&-C_{L,0,2}^{\left( -\right) }C_{L,1,3}^{\left( -\right) }A_{j}^{\dag
}A_{k}A_{l}\rho _{S}^{\left( n\right) }\left( t\right) d_{i}^{\dagger
}-C_{L,0,2}^{\left( -\right) }C_{R,1,3}^{\left( -\right) }A_{j}^{\dag
}A_{k}A_{l}\rho _{S}^{\left( n\right) }\left( t\right) d_{i}^{\dagger }
\notag \\
&&+C_{L,0,2}^{\left( -\right) }C_{L,1,3}^{\left( +\right)
}A_{k}A_{j}A_{l}^{\dag }\rho _{S}^{\left( n\right) }\left( t\right)
d_{i}^{\dagger }+C_{L,0,2}^{\left( -\right) }C_{R,1,3}^{\left( +\right)
}A_{k}A_{j}A_{l}^{\dag }\rho _{S}^{\left( n\right) }\left( t\right)
d_{i}^{\dagger }  \notag \\
&&-C_{L,0,2}^{\left( -\right) }C_{L,1,3}^{\left( +\right)
}A_{j}A_{k}A_{l}^{\dag }\rho _{S}^{\left( n\right) }\left( t\right)
d_{i}^{\dagger }-C_{L,0,2}^{\left( -\right) }C_{R,1,3}^{\left( +\right)
}A_{j}A_{k}A_{l}^{\dag }\rho _{S}^{\left( n\right) }\left( t\right)
d_{i}^{\dagger }  \notag \\
&&+C_{L,0,2}^{\left( +\right) }C_{L,1,3}^{\left( -\right) }A_{k}^{\dag
}A_{j}^{\dag }A_{l}\rho _{S}^{\left( n\right) }\left( t\right)
d_{i}+C_{L,0,2}^{\left( +\right) }C_{R,1,3}^{\left( -\right) }A_{k}^{\dag
}A_{j}^{\dag }A_{l}\rho _{S}^{\left( n\right) }\left( t\right) d_{i}  \notag
\\
&&-C_{L,0,2}^{\left( +\right) }C_{L,1,3}^{\left( -\right) }A_{j}^{\dag
}A_{k}^{\dag }A_{l}\rho _{S}^{\left( n\right) }\left( t\right)
d_{i}-C_{L,0,2}^{\left( +\right) }C_{R,1,3}^{\left( -\right) }A_{j}^{\dag
}A_{k}^{\dag }A_{l}\rho _{S}^{\left( n\right) }\left( t\right) d_{i}  \notag
\\
&&+C_{L,0,2}^{\left( +\right) }C_{L,1,3}^{\left( +\right) }A_{k}^{\dag
}A_{j}A_{l}^{\dag }\rho _{S}^{\left( n\right) }\left( t\right)
d_{i}+C_{L,0,2}^{\left( +\right) }C_{R,1,3}^{\left( +\right) }A_{k}^{\dag
}A_{j}A_{l}^{\dag }\rho _{S}^{\left( n\right) }\left( t\right) d_{i}  \notag
\\
&&-C_{L,0,2}^{\left( +\right) }C_{L,1,3}^{\left( +\right) }A_{j}A_{k}^{\dag
}A_{l}^{\dag }\rho _{S}^{\left( n\right) }\left( t\right)
d_{i}-C_{L,0,2}^{\left( +\right) }C_{R,1,3}^{\left( +\right)
}A_{j}A_{k}^{\dag }A_{l}^{\dag }\rho _{S}^{\left( n\right) }\left( t\right)
d_{i}  \label{I01}
\end{eqnarray}%
\begin{eqnarray}
&&I_{1,n+1}=  \notag \\
&&+C_{R,0,2}^{\left( +\right) }C_{L,1,3}^{\left( -\right) }A_{k}^{\dag
}A_{j}^{\dag }A_{l}\rho _{S}^{\left( n+1\right) }\left( t\right)
d_{i}+C_{R,0,2}^{\left( +\right) }C_{R,1,3}^{\left( -\right) }A_{k}^{\dag
}A_{j}^{\dag }A_{l}\rho _{S}^{\left( n+1\right) }\left( t\right) d_{i}
\notag \\
&&-C_{R,0,2}^{\left( +\right) }C_{L,1,3}^{\left( -\right) }A_{j}^{\dag
}A_{k}^{\dag }A_{l}\rho _{S}^{\left( n+1\right) }\left( t\right)
d_{i}-C_{R,0,2}^{\left( +\right) }C_{R,1,3}^{\left( -\right) }A_{j}^{\dag
}A_{k}^{\dag }A_{l}\rho _{S}^{\left( n+1\right) }\left( t\right) d_{i}
\notag \\
&&+C_{R,0,2}^{\left( +\right) }C_{L,1,3}^{\left( +\right) }A_{k}^{\dag
}A_{j}A_{l}^{\dag }\rho _{S}^{\left( n+1\right) }\left( t\right)
d_{i}+C_{R,0,2}^{\left( +\right) }C_{R,1,3}^{\left( +\right) }A_{k}^{\dag
}A_{j}A_{l}^{\dag }\rho _{S}^{\left( n+1\right) }\left( t\right) d_{i}
\notag \\
&&-C_{R,0,2}^{\left( +\right) }C_{L,1,3}^{\left( +\right) }A_{j}A_{k}^{\dag
}A_{l}^{\dag }\rho _{S}^{\left( n+1\right) }\left( t\right)
d_{i}-C_{R,0,2}^{\left( +\right) }C_{R,1,3}^{\left( +\right)
}A_{j}A_{k}^{\dag }A_{l}^{\dag }\rho _{S}^{\left( n+1\right) }\left(
t\right) d_{i}  \label{I03}
\end{eqnarray}%
where
\begin{equation}
A_{j}=e^{-i\mathcal{L}_{S}\left( t-t_{1}\right) }d_{j},A_{j}^{\dag }=e^{-i%
\mathcal{L}_{S}\left( t-t_{1}\right) }d_{j}^{\dagger },  \label{ajaddsub}
\end{equation}%
\begin{equation}
A_{k}=e^{-i\mathcal{L}_{S}\left( t-t_{2}\right) }d_{k},A_{k}^{\dag }=e^{-i%
\mathcal{L}_{S}\left( t-t_{2}\right) }d_{k}^{\dagger },  \label{akaddsub}
\end{equation}%
\begin{equation}
A_{l}=e^{-i\mathcal{L}_{S}\left( t-t_{3}\right) }d_{l},A_{l}^{\dag }=e^{-i%
\mathcal{L}_{S}\left( t-t_{3}\right) }d_{l}^{\dagger }.  \label{aladdsub}
\end{equation}%
According to the procedure described above, one can obtain the
particle-number-resolved density matrices corresponding to Eq. (\ref{four}),
which is the starting point of the non-Markovian FCS calculation taking the
cotunneling processes into account. Therefore, the particle number resolved
quantum master equation taking into account both the sequential tunneling
and cotunneling can be written as%
\begin{equation}
\frac{\partial \rho _{S}^{\left( n\right) }\left( t\right) }{\partial t}=-i%
\left[ H_{S},\rho _{S}\left( t\right) \right] +\left. \frac{\partial \rho
_{S}^{\left( n\right) }\left( t\right) }{\partial t}\right\vert _{\text{%
sceond-order}}+\left. \frac{\partial \rho _{S}^{\left( n\right) }\left(
t\right) }{\partial t}\right\vert _{\text{fourth-order}}  \label{twoandfour}
\end{equation}

\subsection{FULL COUNTING STATISTICS}

The FCS formalism based on Eq. (\ref{twoandfour}) can be obtained from the
cumulant generating function (CGF) $F\left( \chi \right) $ \cite{Bagrets}%
\begin{equation}
e^{-F\left( \chi \right) }=\sum_{n}P\left( n,t\right) e^{in\chi },
\label{CGF}
\end{equation}%
where $\chi $ is the counting field, and $P\left( n,t\right) =$Tr$\left[
\rho _{S}^{\left( n\right) }\left( t\right) \right] $. Thus, one has $%
e^{-F\left( \chi \right) }=$Tr$\left[ S\left( \chi ,t\right) \right] $ by
defining $S\left( \chi ,t\right) =\sum_{n}\rho _{S}^{\left( n\right) }\left(
t\right) e^{in\chi }$, where the trace is over the eigenstates of the OQS.
Since Eq. (\ref{twoandfour}) has the following form
\begin{equation}
\dot{\rho}_{S}^{\left( n\right) }=A\rho _{S}^{\left( n\right) }+C_{1}\rho
_{S}^{\left( n+1\right) }+D_{1}\rho _{S}^{\left( n-1\right) }+C_{2}\rho
_{S}^{\left( n+2\right) }+D_{2}\rho _{S}^{\left( n-2\right) },
\label{formalmaster}
\end{equation}%
then $S\left( \chi ,t\right) $ satisfies
\begin{equation}
\dot{S}=AS+e^{-i\chi }C_{1}S+e^{i\chi }D_{1}S+e^{-2i\chi }C_{2}S+e^{2i\chi
}D_{2}S\equiv L_{\chi }S,  \label{formalmaster1}
\end{equation}%
where $S$ is a column matrix, and $A$, $C_{1}$, $D_{1}$, $C_{2}$ and $D_{2}$
are five square matrices. Here, for the second-order case $C_{2}=D_{2}=0$,
and the specific form of $L_{\chi }$ can be obtained by performing a
discrete Fourier transformation to the matrix elements of Eq. (\ref%
{twoandfour}).

In the low frequency limit, the low order cumulants of
transferred-electron number $C_{k}$ can be calculated based on Eq. (\ref%
{formalmaster1}) and the Rayleigh--Schr\"{o}dinger perturbation theory
developed in Refs. \cite%
{Xueoff3,Flindt01,Flindt02,Flindt03,WangSK,Bagrets,Kielich,Groth}. Here, the
first four cumulants are directly related to the peak position (i.e., the
average current $\left\langle I\right\rangle =eC_{1}/t$), the peak-width
(i.e., shot noise characterized by Fano factor $C_{2}/C_{1}$), the skewness (%
$C_{3}/C_{1}$) and the kurtosis ($C_{4}/C_{1}$) of the distribution of
transferred-electron number. In general, the shot noise, skewness and
kurtosis are represented by the Fano factors $F_{2}=C_{2}/C_{1}$, $%
F_{3}=C_{3}/C_{1}$ and $F_{4}=C_{4}/C_{1}$, respectively.

\section{TRANSPORT THROUGH SIDE-COUPLED DOUBLE QD SYSTEM}

\subsection{Hamiltonian of the side-coupled double QD system}

In order to facilitate discussions effectively, we consider a side-coupled
double QD system weakly connected to two metallic electrodes, see Fig. 1.
For the sake of simplicity, we neglect electron-spin. The Hamiltonian of the
side-coupled double-QD system is described by%
\begin{equation}
H_{\text{dot}}=\varepsilon _{1}d_{1}^{\dag }d_{1}+\varepsilon
_{2}d_{2}^{\dag }d_{2}+U_{12}\hat{n}_{1}\hat{n}_{2}-J\,\left( d_{1}^{\dag
}d_{2}+d_{2}^{\dag }d_{1}\right) ,  \label{sidecoupled}
\end{equation}%
where $d_{i}^{\dag }$ ($d_{i}$) is the creation (annihilation) operator of
an electron with energy $\varepsilon _{i}$ in $i$th QD, and $U_{12}$ the
interdot Coulomb repulsion between two electrons in different QDs. Here, we
assume that the intradot Coulomb interaction $U\rightarrow \infty $, thus,
the double-electron occupation in different QDs is permitted only. The last
term of $H_{dot}$ describes the hopping between the two QDs with $J$ being
the hopping parameter. To facilitate the following calculation, the
eigenstates of $H_{\text{dot}}$ are used to describe the electronic states
of the side-coupled double QD system. Here, The Hamiltonian $H_{\text{dot}}$
can be diagonalized in the basis represented by the electron occupation
numbers of the QD-1 and the QD-2, i.e., $\left\vert 0\right\rangle
_{1}\left\vert 0\right\rangle _{2}$, $\left\vert 1\right\rangle
_{1}\left\vert 0\right\rangle _{2}$, $\left\vert 0\right\rangle
_{1}\left\vert 1\right\rangle _{2}$, $\left\vert 1\right\rangle
_{1}\left\vert 1\right\rangle _{2}$. Consequently, the four eigenvalues of
and the corresponding four eigenstates of the side-coupled double QD system
are given by \cite{Xueoff3}
\begin{equation}
H_{\text{dot}}\left\vert 0\right\rangle =0,\left\vert 0\right\rangle
=\left\vert 0\right\rangle _{1}\left\vert 0\right\rangle _{2},
\label{eigenstates0}
\end{equation}%
\begin{equation}
H_{\text{dot}}\left\vert 1\right\rangle ^{\pm }=\varepsilon _{\pm
}\left\vert 1\right\rangle ^{\pm },\left\vert 1\right\rangle ^{\pm }=a_{\pm
}\left\vert 1\right\rangle _{1}\left\vert 0\right\rangle _{2}+b_{\pm
}\left\vert 0\right\rangle _{1}\left\vert 1\right\rangle _{2},
\label{eigenstates1}
\end{equation}%
\begin{equation}
H_{\text{dot}}\left\vert 2\right\rangle =\varepsilon _{1,1}\left\vert
2\right\rangle ,\left\vert 2\right\rangle =\left\vert 1\right\rangle
_{1}\left\vert 1\right\rangle _{2},  \label{eigenstates2}
\end{equation}%
with
\begin{equation}
\varepsilon _{\pm }=\frac{\left( \varepsilon _{1}+\varepsilon _{2}\right)
\pm \sqrt{\left( \varepsilon _{1}-\varepsilon _{2}\right) ^{2}+4J^{2}}}{2},
\label{eigenvalues1}
\end{equation}%
\begin{equation}
\varepsilon _{1,1}=\varepsilon _{1}+\varepsilon _{2}+U_{12}
\label{eigenvalues2}
\end{equation}%
and%
\begin{equation}
a_{\pm }=\frac{\mp J}{\sqrt{\left( \varepsilon _{\pm }-\varepsilon
_{1}\right) ^{2}+J^{2}}},  \label{abaddsub}
\end{equation}%
\begin{equation}
b_{\pm }=\frac{\pm \left( \varepsilon _{\pm }-\varepsilon _{1}\right) }{%
\sqrt{\left( \varepsilon _{\pm }-\varepsilon _{1}\right) ^{2}+J^{2}}}.
\label{bbaddsub}
\end{equation}

The electron distributions of the two metallic electrodes, in which the
relaxation is assumed to be sufficiently fast, are described by the
equilibrium Fermi functions and the corresponding Hamiltonian reads%
\begin{equation}
H_{\text{electrodes}}=\sum_{\alpha \mathbf{k}}\varepsilon _{\alpha \mathbf{k}%
}a_{\alpha \mathbf{k}}^{\dag }a_{\alpha \mathbf{k}}  \label{QDelectrodes}
\end{equation}
where $a_{\alpha \mathbf{k}}^{\dag }$ ($a_{\alpha \mathbf{k}}$) is the $%
\alpha$-electrode electron creation (annihilation) operator with energy $%
\varepsilon _{\alpha \mathbf{k}}$ and momentum $\mathbf{k}$.

The tunneling between the QD-1 and the two electrodes is described by%
\begin{equation}
H_{\text{hyb}}=\sum_{\alpha \mathbf{k}}\left( t_{\alpha \mathbf{k}}a_{\alpha
\mathbf{k}}^{\dag }d_{1}+t_{\alpha \mathbf{k}}^{\ast }d_{1}^{\dag }a_{\alpha
\mathbf{k}}\right) ,  \label{tunneling}
\end{equation}
where the tunneling amplitudes $t_{\alpha }$ and the density of states $%
g_{\alpha }$ are assumed to be independent of wave vector and energy, thus,
the electronic tunneling rate can be characterized by $\Gamma _{\alpha
}=2\pi |t_{\alpha }|^{2}g_{\alpha }$.

In the side-coupled double QD system, the quantum coherence can be tuned by
modulating the magnitude of the hopping parameter $J$ relative to the tunneling
coupling strength between the QD-1 and the two electrodes. In the case of $%
J\ll \Gamma $ ($\Gamma =\Gamma _{L}+\Gamma _{R}$), the hopping strength
between the two QDs strongly modifies the internal dynamics, and the
off-diagonal elements of the reduced density matrix play an essential role
in the electron tunneling processes \cite{Xueoff1,Xueoff3,LuoJPCM}; while in
the regime $J\gg \Gamma $, the off-diagonal elements of the reduced density
matrix have very little influence on the electron tunneling processes \cite%
{Xueoff1}. In the following calculation, the parameters of the side-coupled
double QD system are taken as $\varepsilon _{1}=\varepsilon _{2}=2.35$, $%
U_{12}=4$ and $k_{B}T=0.1$ (if not explicitly stated otherwise), where the
unit of energy is chosen as meV \cite{Elzerman}.

\subsection{The side-coupled double QD system with strong quantum coherence}

We first study the influences of the off-diagonal elements of the reduced
density matrix, namely, the quantum coherence, and the electron cotunneling
processes on the FCS of electron transport through this QD system with
strong quantum coherence. For the side-coupled double QD system, the quantum
coherence has an important influence on the electron sequential tunneling
processes in the bias voltage range in which the transitions between the
singly-occupied and empty-occupied eigenstates take place \cite%
{Xueoff1,Xueoff3}. Consequently, the following discussions focus on this
bias voltage region, the applied bias voltage is here chosen as $V_{b}=4.5$
based on the parameters of the QD system. To determine the dependence of the
FCS on the quantum coherence and the electron cotunneling processes, we
consider the average current, shot noise, skewness and kurtosis as a
function of the tunneling rate $\Gamma _{\alpha }$ for the four different
cases, (1) considering the diagonal elements of the reduced density matrix
in the sequential tunneling processes only, (2) considering the diagonal and
off-diagonal elements of the reduced density matrix in the sequential
tunneling processes, (3) considering the diagonal elements of the reduced
density matrix in the cotunneling assisted sequential tunneling processes
only, (4) considering the diagonal and off-diagonal elements of the reduced
density matrix in\ the cotunneling assisted sequential tunneling processes.

Figures 2, 3 and 4 show the influence of the temperature of the QD system on
the first four current cumulants with the different values of the left-right
asymmetry of the QD-electrode coupling $\Gamma _{L}/\Gamma _{R}$. In the
case of the coupling of the QD-1 with the source-electrode is stronger than
that of the QD-1 with the drain-electrode, i.e., $\Gamma _{L}/\Gamma _{R}>1$%
, we in Fig. 2 plot the first four current cumulants as a function of the
tunneling rate $\Gamma _{L}$ with different temperatures $k_{B}T$ at $\Gamma
_{L}/\Gamma _{R}=10$. We found that, in the $\Gamma /J<1$ case, the electron
cotunneling processes play a essential role in determining the values of the
shot noise and high-order current cumulants; whereas in the $\Gamma /J\gg 1$
case the quantum coherence plays a crucial role in determining whether the
Fano factors of the shot noise, the skewness and the kurtosis are larger
than one or not, see Fig. 2. In the case of the intermediate value of $%
\Gamma /J$, the competition between the electron cotunneling processes and
the quantum coherence takes place. This leads to the formation of a
crossover region, but the range of which depends on the temperature $k_{B}T$%
, see Fig. 2.

The underlying physics of the cotunneling effect can be understood in terms
of the cotunneling-induced redistribution of the occupation probabilities
of the QD's different eigenstates. In the $\Gamma _{L}/\Gamma _{R}=10$ case, the occupation
probabilities of the two singly-occupied eigenstates are much larger than
that of empty-occupied eigenstate, leading to a relatively long dwell time
of the conduction electron before tunneling out the QD system. When $\Gamma
/J\ll 1$, the conduction electrons can tunnel back and forth between the two
singly-occupied eigenstates very rapidly, and then enhance the cotuneling
processes induced by the transitions between the doubly-occupied $\left\vert
2\right\rangle $ and singly-occupied $\left\vert 1\right\rangle ^{\pm }$
eigenstates. Consequently, the cotunneling processes can dramatically
decrease and increase the occupation probabilities of the singly-occupied
and empty-occupied eigenstates with decreasing ratio of $\Gamma $ to $J$,
respectively, see Figs. 5(a1)-5(a3). This indicates that the sequential-induced
blocking of electron tunneling can be removed by the cotunneling processes,
which leads to the shot noise being decreased, see Figs. 2(b1)-2(b3).
Whereas in the $\Gamma /J\gg 1$ case the conduction electrons can tunnel back and forth
between the two singly-occupied eigenstates very slowly, and then suppress the
cotuneling processes. Thus, the quantum coherence has a very significant influence on
the electron tunneling processes and the cotunneling-induced probability
distributions for the singly-occupied and empty-occupied eigenstates have a
slight variation, see Figs. 5(a1)-5(a3). In addition, the cotunneling-induced
non-equilibrium electron distribution depends on the temperature $k_{B}T$,
see Figs. 5(a1)-5(a3), which are responsible for the slight influence of the
cotunneling effect on the FCS with decreasing the temperature $k_{B}T$.

Compared with the $\Gamma _{L}/\Gamma _{R}>1$ case, in the case of $\Gamma
_{L}/\Gamma _{R}\leq 1$, the range of the crossover region is very small,
which also depends on the temperature $k_{B}T$, see Figs. 3 and 4.
Particularly, in the cases of $\Gamma _{L}/\Gamma _{R}\leq 1$ and $\Gamma
/J\gg 1$, the interplay between the electron cotunneling processes and the
quantum coherence determine the FCS properties of transferred-electron
number, such as, whether the super-Poissonian distributions of the shot
noise, the skewness and the kurtosis ($F_{i}>1$) occur or not, and whether
the signs of the values of the skewness and the kurtosis become negative
from a positive value or not, see Figs. 3 and 4. In particular,
the magnitudes and signs of the skewness and kurtosis characterize the
asymmetry of and the combined weight of the tails relative to the rest of
the probability distribution of transferred-electron number, respectively.
Thus, they can provide much more information for the counting
statistics that the shot noise. Moreover, in the cases of $\Gamma _{L}/\Gamma _{R}=1$ and $\Gamma /J\gg 1$,
the behavior of the shot noise is mainly governed by the quantum coherence,
see Figs. 4(b1), 4(b2) and 4(b3). However, these characteristics also depend
on the temperature $k_{B}T$, i.e., the quantum coherence will play an
essential role in the electron tunneling processes with decreasing
temperature, see Figs. 3(a3)-3(d3) and 4(a3)-4(d3).

These properties of the $\Gamma_{L}/\Gamma _{R}\leq 1$ case can also be explained
through the cotunneling-induced redistribution of the occupation probabilities.
In the $\Gamma _{L}/\Gamma _{R}=0.1$ case, the occupation probabilities of
the two singly-occupied eigenstates are much smaller than that of
empty-occupied eigenstate, thus, the conduction electrons have a very short
dwell time, which is contrary to the $\Gamma _{L}/\Gamma _{R}=10$ case.
In the $\Gamma /J\gg 1$ case, the two electron tunneling
can occur through the cotunneling processes induced by the transitions
between the doubly-occupied $\left\vert 2\right\rangle $ and singly-occupied
$\left\vert 1\right\rangle ^{\pm }$ eigenstates and the succeed sequential
processes induced by the transitions between the singly-occupied $\left\vert
1\right\rangle ^{\pm }$ and empty-occupied $\left\vert 0\right\rangle $
eigenstates. Thus, the cotunneling assisted sequential tunneling processes
in the $\Gamma /J\gg 1$ case still play a important role in determining the
FCS. Additionally, in the $\Gamma /J\ll 1$ case, the cotunneling processes can further increase the
occupation probability of the empty-occupied eigenstate with decreasing
ratio of $\Gamma $ to $J$, see Figs. 5(b1)-5(b3). This effect can further block the
electron tunneling, then leading to the shot noise being relatively
enhanced, see Figs. 3(b1)-3(b3).

Figures 6 and 7 show the influence of the left-right asymmetry of the
QD-electrode coupling $\Gamma _{L}/\Gamma _{R}$ on the first four current
cumulants for a given temperature $k_{B}T=0.1$. In the cases of $\Gamma
_{L}/\Gamma _{R}>1$ and $\Gamma >J$ (the intermediate value), the range of
the crossover region, in which the electron cotunneling processes decrease
the values of Fano factors while the quantum coherence increase that of Fano
factors, increases with increasing the ratio of $\Gamma _{L}$ to $\Gamma _{R}
$, see Fig. 6. Whereas in the cases of $\Gamma _{L}/\Gamma _{R}<1$ and $%
\Gamma /J\gg 1$, the interplay between the electron cotunneling processes
and the quantum coherence has a relatively remarkable influence on the
high-order current cumulants with decreasing the ratio of $\Gamma _{L}$ to $%
\Gamma _{R}$, and determines whether the super-Poissonian distributions of
the shot noise and the kurtosis take place or not, and whether the
transition of the skewness from positive to negative values occurs or not,
see Fig. 7. These results can also be understood with the help of the
redistribution of the occupation probability induced by the left-right asymmetry of the
QD-electrode coupling, see Fig. 8.

\subsection{The side-coupled double QD system with weak quantum coherence}

We finally discuss the influences of the electron cotunneling processes on
the first four current cumulants in the side-coupled double QD system with
weak quantum coherence. Here, the hopping parameter is thus chosen as $J=1$.
According to the parameters of the QD system, we choose the three fixed bias
voltages, under which the different transitions between the QD eigenstates
participate in the electron tunneling processes, namely, $V_{b}=2.5$
corresponding to the transitions between the singly-occupied $\left\vert
1\right\rangle ^{-}$ and empty-occupied eigenstates, $V_{b}=4.5$
corresponding to the transitions between the singly-occupied $\left\vert
1\right\rangle ^{\pm }$ and empty-occupied eigenstates, and $V_{b}=6.5$
corresponding to the transitions between the singly-occupied $\left\vert
1\right\rangle ^{\pm }$ and empty-occupied eigenstates and the transitions
between the doubly-occupied $\left\vert 2\right\rangle $ and singly-occupied
$\left\vert 1\right\rangle ^{+}$ eigenstates. In this situation, the
properties of the first four current cumulants are well determined by the
electron cotunneling processes because the quantum coherence indeed has a
very small influence on the values of the first four current cumulants, see
Figs. 9 and 10. In the case of $\Gamma _{L}/\Gamma _{R}>1$, the electron
cotunneling processes have a relatively obvious influence on the first four
current cumulants, see Fig. 9; whereas in the case of $\Gamma
_{L}/\Gamma _{R}<1$ that have a slight influence on the first four current
cumulants, see Fig. 10. It is important to note that the electron
cotunneling processes do not change the intrinsic statistical properties of
current cumulants, namely, whether the super-Poissonian distribution of the
current cumulants take place or not, and whether the sign transitions of the
values of the skewness and the kurtosis occur or not, see Figs. 9 and 10.

\section{CONCLUSIONS}

We have developed an efficient non-Markovian FCS formalism taking into
account both the sequential tunneling and cotunneling processes, and studied
the influences of the quantum coherence and the cotunneling assisted
sequential tunneling processes on the first four current cumulants in a
side-coupled double QD system. In the strong quantum-coherent side-coupled
double QD system, it is numerically demonstrated that, in the sequential
tunneling regime, the competition or interplay between the cotunneling
processes and the quantum coherence determines that whether the
super-Poissonian distributions of the shot noise, the skewness and the
kurtosis take place, and whether the sign transitions of the values of the
skewness and the kurtosis occur. These characteristics depend on the
temperature of the QD system, the left-right asymmetry of the QD-electrode
coupling, and the magnitude of the coupling strengths. However, in the weak
quantum-coherent side-coupled double QD system, the cotunneling processes
has a relatively slight influence on the statistical properties of current
cumulants, which also depends on the left-right asymmetry of the
QD-electrode coupling and the corresponding coupling strengths.
Consequently, the dependence of the FCS on the quantum coherence and the
cotunneling processes is necessary to be considered in the open strong
quantum-coherent quantum systems, even through the electron tunneling is
mainly dominated by the sequential tunneling processes.

\section{ACKNOWLEDGMENTS}

This work was supported by the Shanxi Natural Science Foundation of China
under Grant No. 201601D011015, Program for the Outstanding Innovative Teams
of Higher Learning Institutions of Shanxi, NKRDP under Grant No.
2016YFA0301500, NSFC under Grants Nos. 11204203, 11434015, 61227902,
61378017, KZ201610005011, SKLQOQOD under Grant No. KF201403, SPRPCAS under
Grants Nos. XDB01020300, XDB21030300.

\newpage

\begin{figure*}[t]
\centerline{\includegraphics[height=12cm,width=16cm]{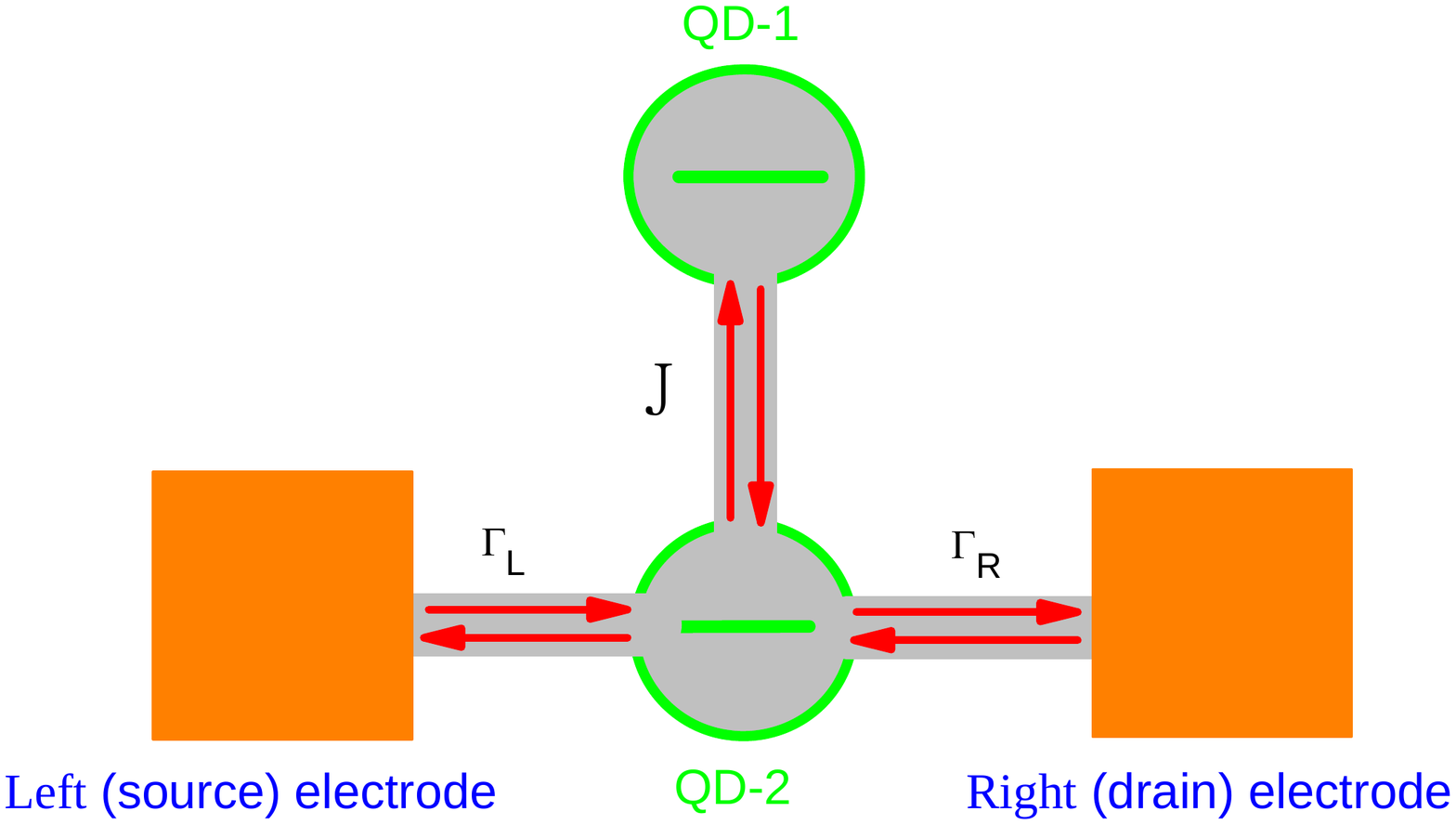}}
\caption{(Color online) The open quantum system consists of a
quantum-coherent-tunable side-coupled single-level double quantum-dot (QD)
system weakly coupled to two electron reservoirs (electrodes). Here, $J$ and
$\Gamma _{\alpha }$ characterize the hopping between the two QDs and the
tunneling coupling between the QD-1 and the electrode $\alpha $,
respectively. The QD molecule possess strong quantum coherence in the case
of $J\ll \Gamma $ ($\Gamma =\Gamma _{L}+\Gamma _{R}$), whereas in the case
of $J\gg \Gamma $ that possess weak quantum coherence.} \label{fig1}
\end{figure*}

\newpage

\begin{figure*}[t]
\centerline{\includegraphics[height=10cm,width=16cm]{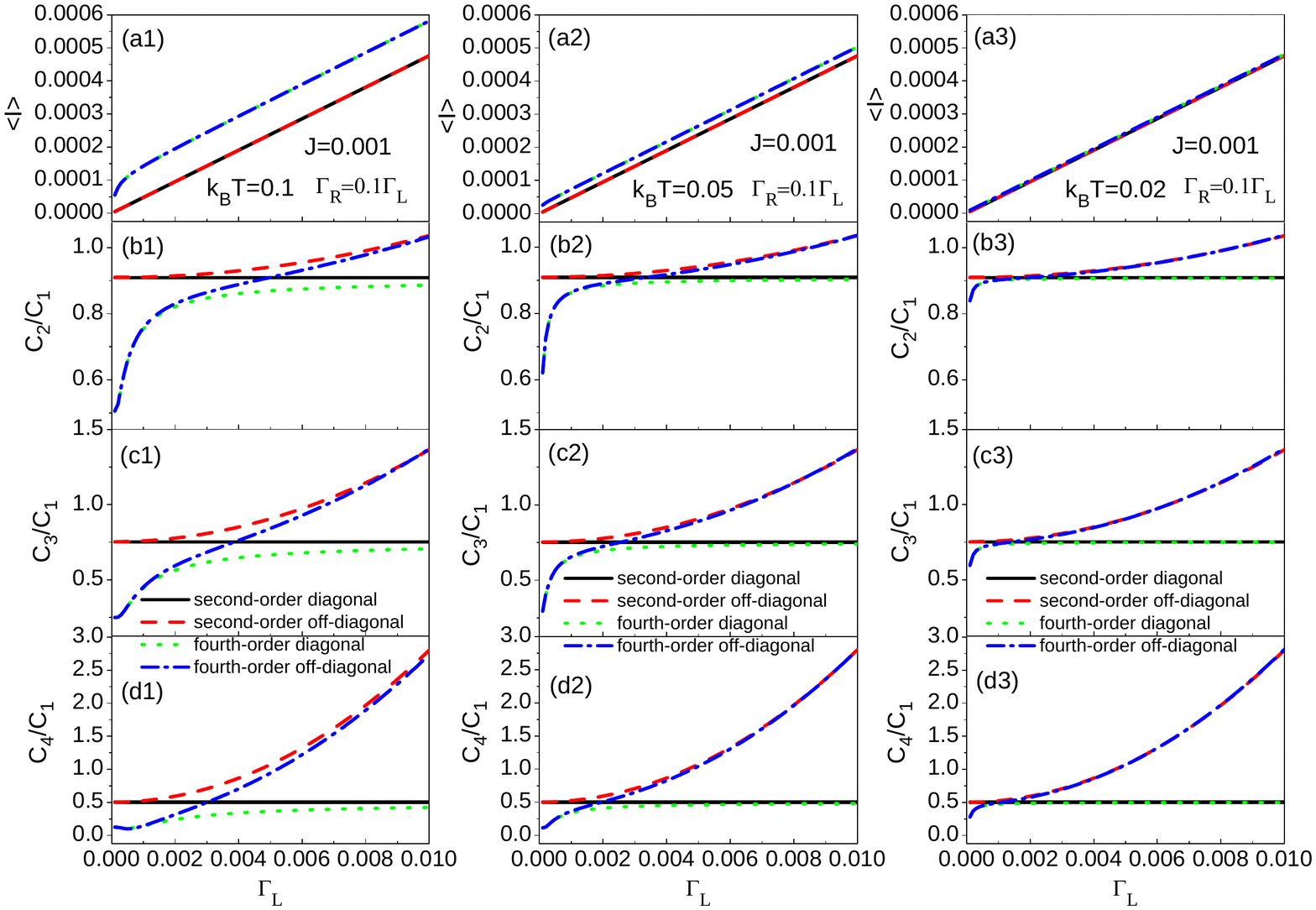}}
\caption{(Color online) The average current $\left\langle I\right\rangle $,
shot noise $C_{2}/C_{1}$, skewness $C_{3}/C_{1}$ and kurtosis $C_{4}/C_{1}$
as a function of the tunneling rate $\Gamma _{L}$ with different
temperatures of the QD system $k_{B}T$ at $\Gamma _{L}/\Gamma _{R}=10$,
where $C_{k}$ is the zero-frequency ${k}$-order cumulant of
transferred-electron number. Here, the four different cases are considered,
namely, (1) considering the diagonal elements of the reduced density
matrix in the sequential tunneling processes only, denoted by second-order
diagonal, (2) considering the diagonal and off-diagonal elements of the
reduced density matrix in the sequential tunneling processes, denoted by
second-order off-diagonal, (3) considering the diagonal elements of the
reduced density matrix in the cotunneling assisted sequential tunneling
processes only, denoted by fourth-order diagonal, (4) considering the diagonal
and off-diagonal elements of the reduced density matrix in the cotunneling
assisted sequential tunneling processes, denoted by fourth-order
off-diagonal. In the case of $\Gamma /J<1$, the properties of current
cumulants are mainly governed by the electron cotunneling processes; whereas
in the case of $\Gamma /J\gg 1$ that are mainly governed by the quantum
coherence. In the case of the intermediate value of $\Gamma /J$, the
competition between the electron cotunneling processes and the quantum
coherence takes place, which leads to the formation of a crossover region.
However, the range of the crossover region depends on the temperature $%
k_{B}T $. The side-coupled double QD system parameters: $\epsilon
_{1}=\epsilon _{2}=2.35$, $J=0.001$, $U_{12}=4$ and $V_{b}=4.5$, where meV
is chosen as the unit of energy.} \label{fig2}
\end{figure*}

\newpage

\begin{figure*}[t]
\centerline{\includegraphics[height=12cm,width=16cm]{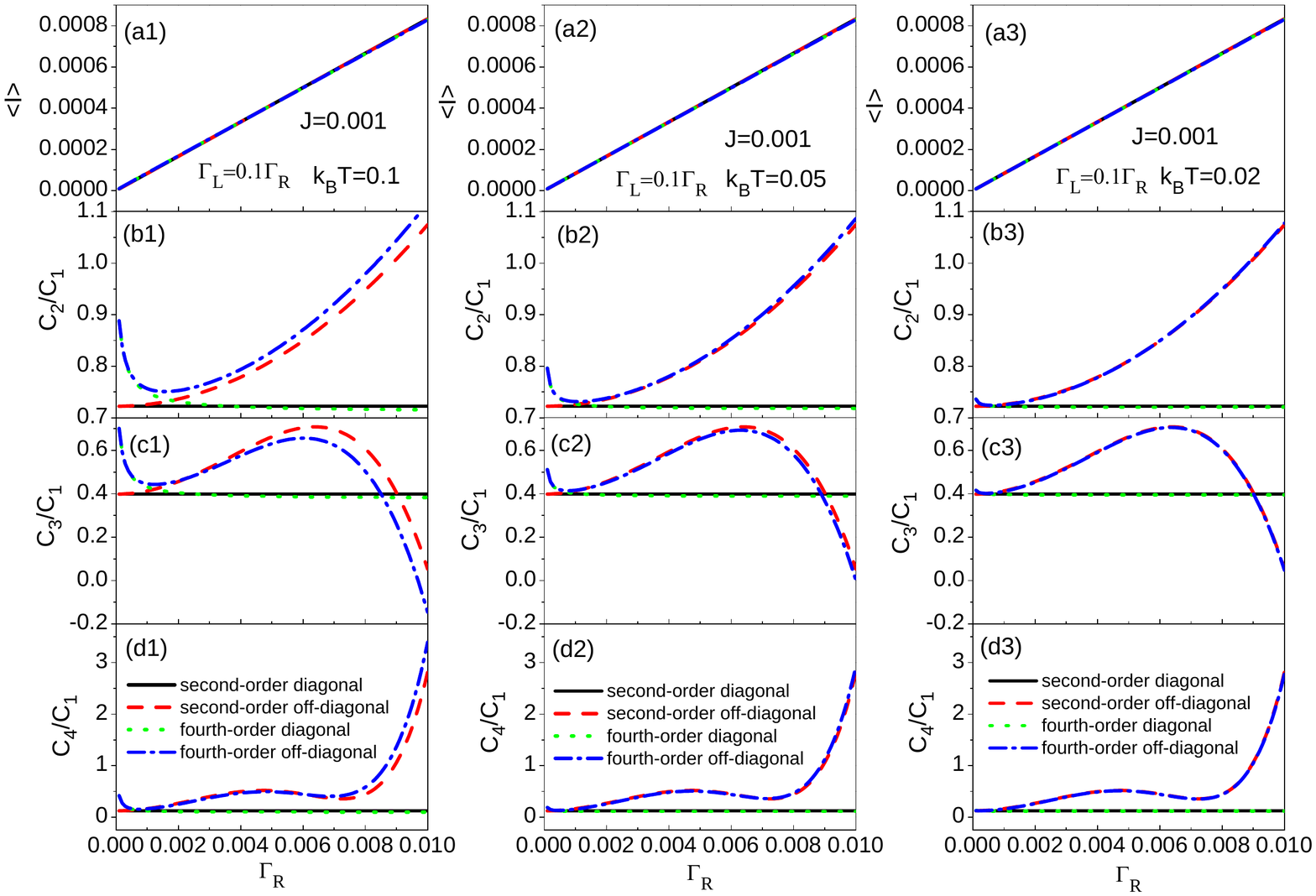}}
\caption{(Color online) The average current $\left\langle I\right\rangle $,
shot noise $C_{2}/C_{1}$, skewness $C_{3}/C_{1}$ and kurtosis $C_{4}/C_{1}$
as a function of the tunneling rate $\Gamma _{R}$ with different
temperatures of the QD system $k_{B}T$ at $\Gamma _{L}/\Gamma _{R}=0.1$. In
the $\Gamma /J\gg 1$ case, the interplay between the electron cotunneling
processes and the quantum coherence determines whether the super-Poissonian
distributions of the shot noise and the kurtosis ($F_{i}>1$) occur, and
whether the signs of the values of the skewness become negative from
positive values, which also depends on the temperature $k_{B}T$. The
notations and the parameters of the QD system are the same as in Fig. 2.} \label{fig3}
\end{figure*}

\newpage

\begin{figure*}[t]
\centerline{\includegraphics[height=12cm,width=16cm]{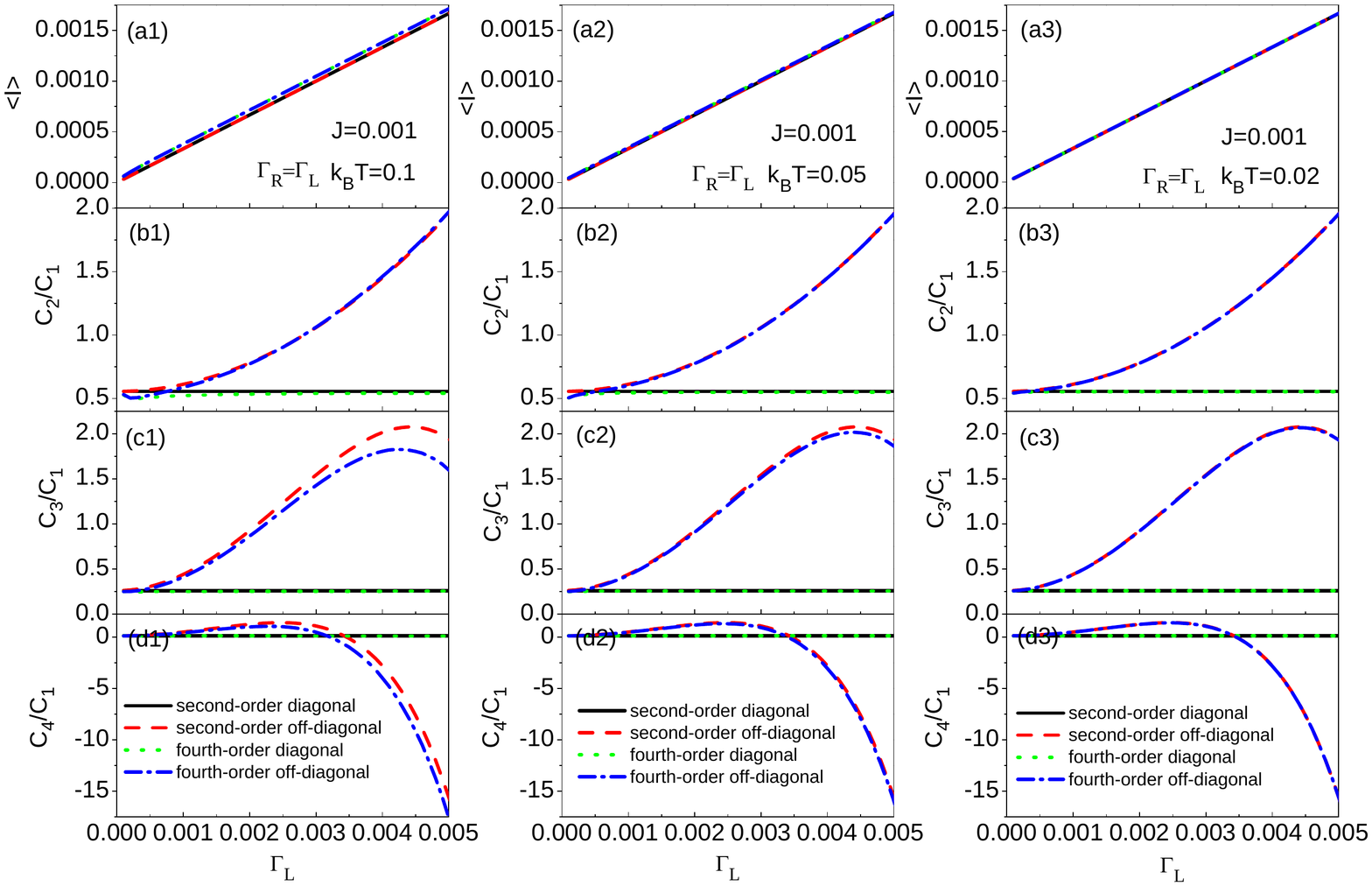}}
\caption{(Color online) The average current $\left\langle I\right\rangle $,
shot noise $C_{2}/C_{1}$, skewness $C_{3}/C_{1}$ and kurtosis $C_{4}/C_{1}$
as a function of the tunneling rate $\Gamma _{L}$ with different
temperatures of the QD system $k_{B}T$ at $\Gamma _{L}/\Gamma _{R}=1$. In
the $\Gamma /J\gg 1$ case, the quantum coherence plays an essential role in
determining whether the super-Poissonian shot noise takes place; whereas the
interplay between the electron cotunneling processes and the quantum
coherence determines whether the super-Poissonian distributions of the
skewness and the kurtosis occur, and whether the signs of the values of the
kurtosis become a large negative from a small positive values, which depends
on the temperature $k_{B}T$. The notations and the parameters of the QD
system are the same as in Fig. 2.} \label{fig4}
\end{figure*}

\newpage

\begin{figure*}[t]
\centerline{\includegraphics[height=12cm,width=16cm]{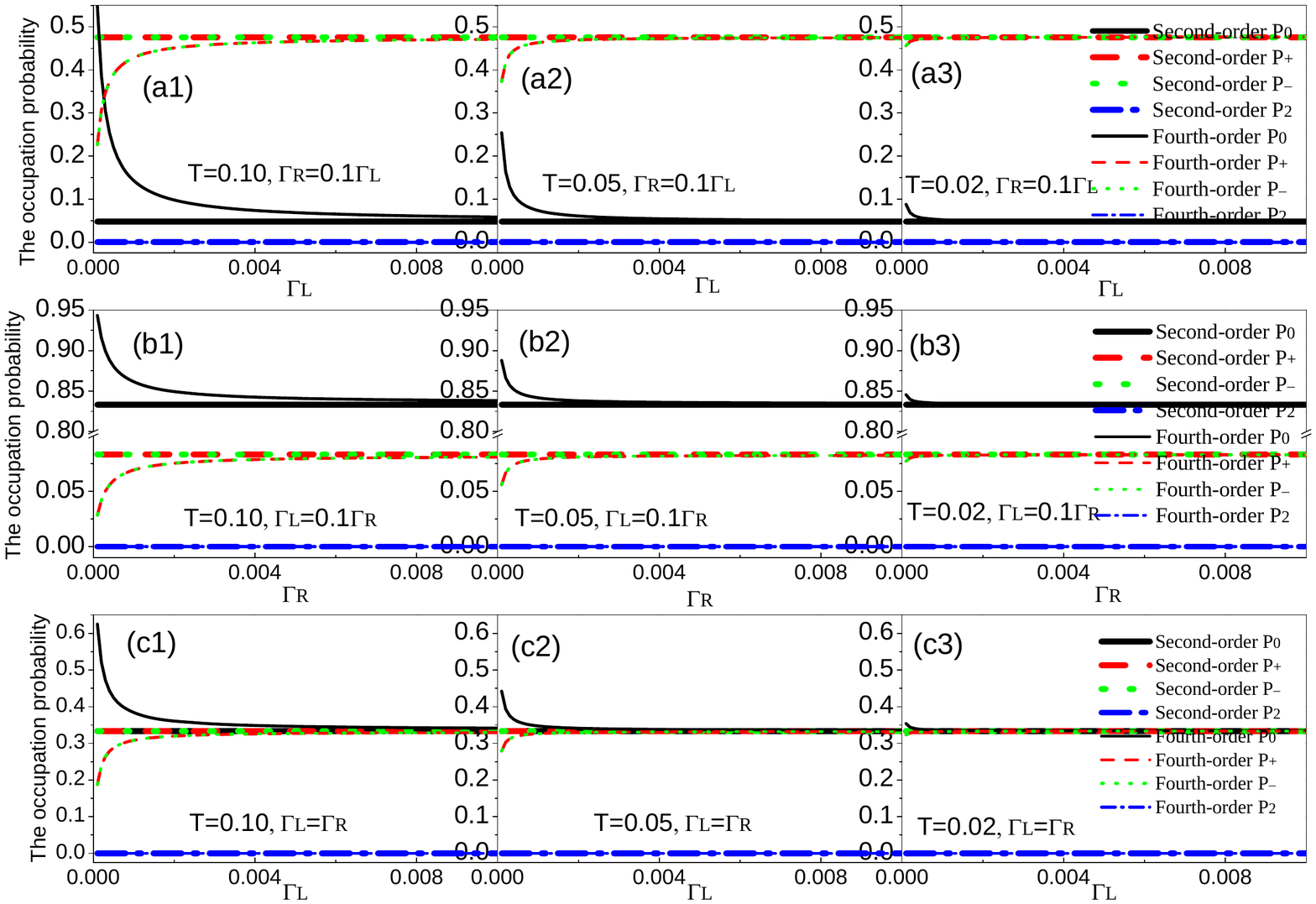}}
\caption{(Color online) The occupation probabilities of the QD's eigenstates
as a function of the tunneling rate $\Gamma _{L}$ ($\Gamma _{R}$) with different values of
the ratio of $\Gamma _{L}$ to $\Gamma _{R}$ and the temperature $k_{B}T$.
The parameters of the QD system of (a1)-(a3), (b1)-(b3) and (c1)-(c3) are
the same as in Figs. 2, 3 and 4, respectively.} \label{fig5}
\end{figure*}

\newpage

\begin{figure*}[t]
\centerline{\includegraphics[height=12cm,width=16cm]{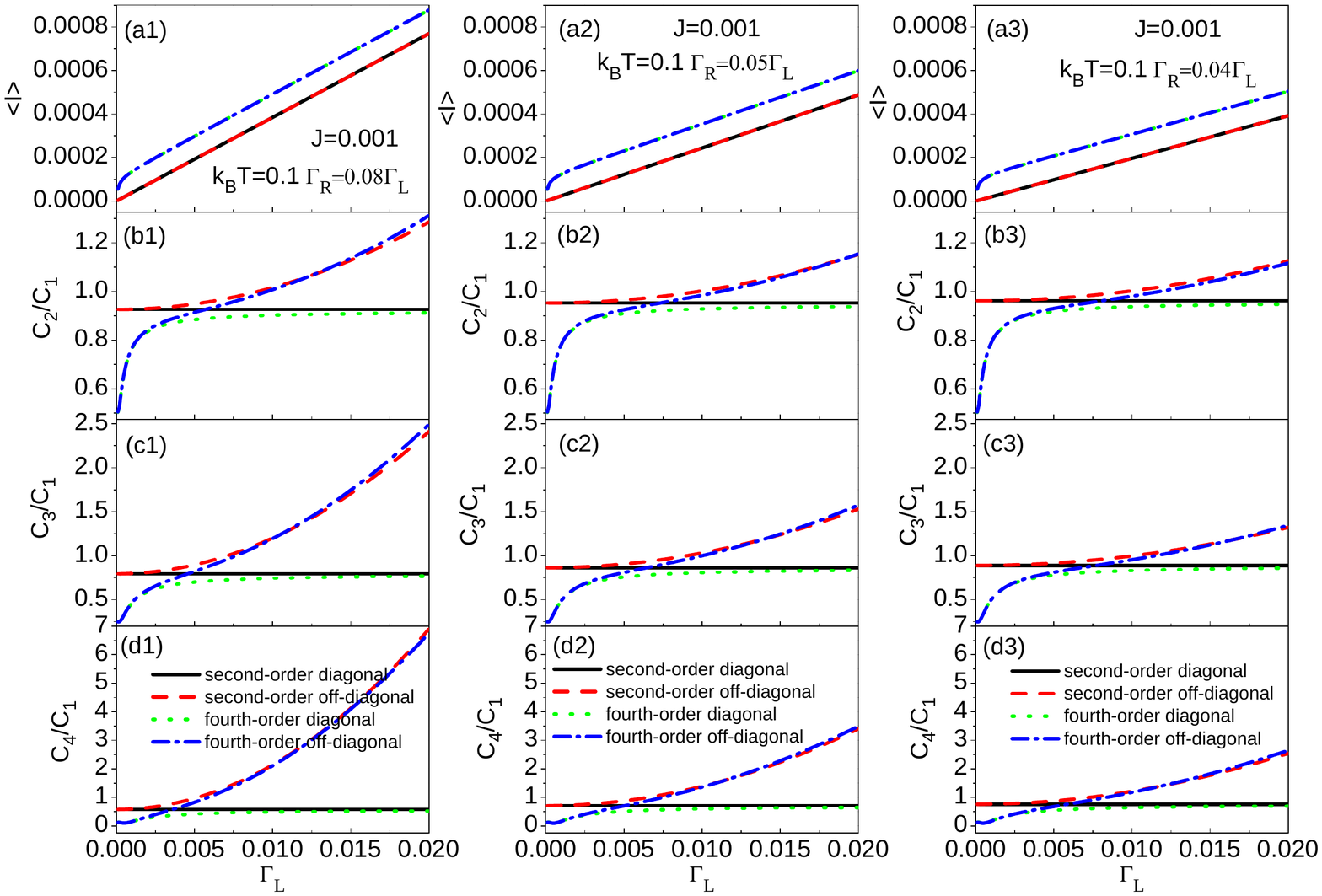}}
\caption{(Color online) The average current $\left\langle I\right\rangle $,
shot noise $C_{2}/C_{1}$, skewness $C_{3}/C_{1}$ and kurtosis $C_{4}/C_{1}$
as a function of the tunneling rate $\Gamma _{L}$ with different values of
the ratio of $\Gamma _{L}$ to $\Gamma _{R}$ at $\Gamma _{L}>\Gamma _{R}$ and
$k_{B}T=0.1$. In the crossover region, the electron cotunneling processes
decrease the values of Fano factors, while the quantum coherence increase
that of Fano factors. However, the range of the crossover region increases
with increasing the ratio of $\Gamma _{L}$ to $\Gamma _{R}$. The notations
and the parameters of the QD system are the same as in Fig. 2.} \label{fig6}
\end{figure*}

\newpage

\begin{figure*}[t]
\centerline{\includegraphics[height=12cm,width=16cm]{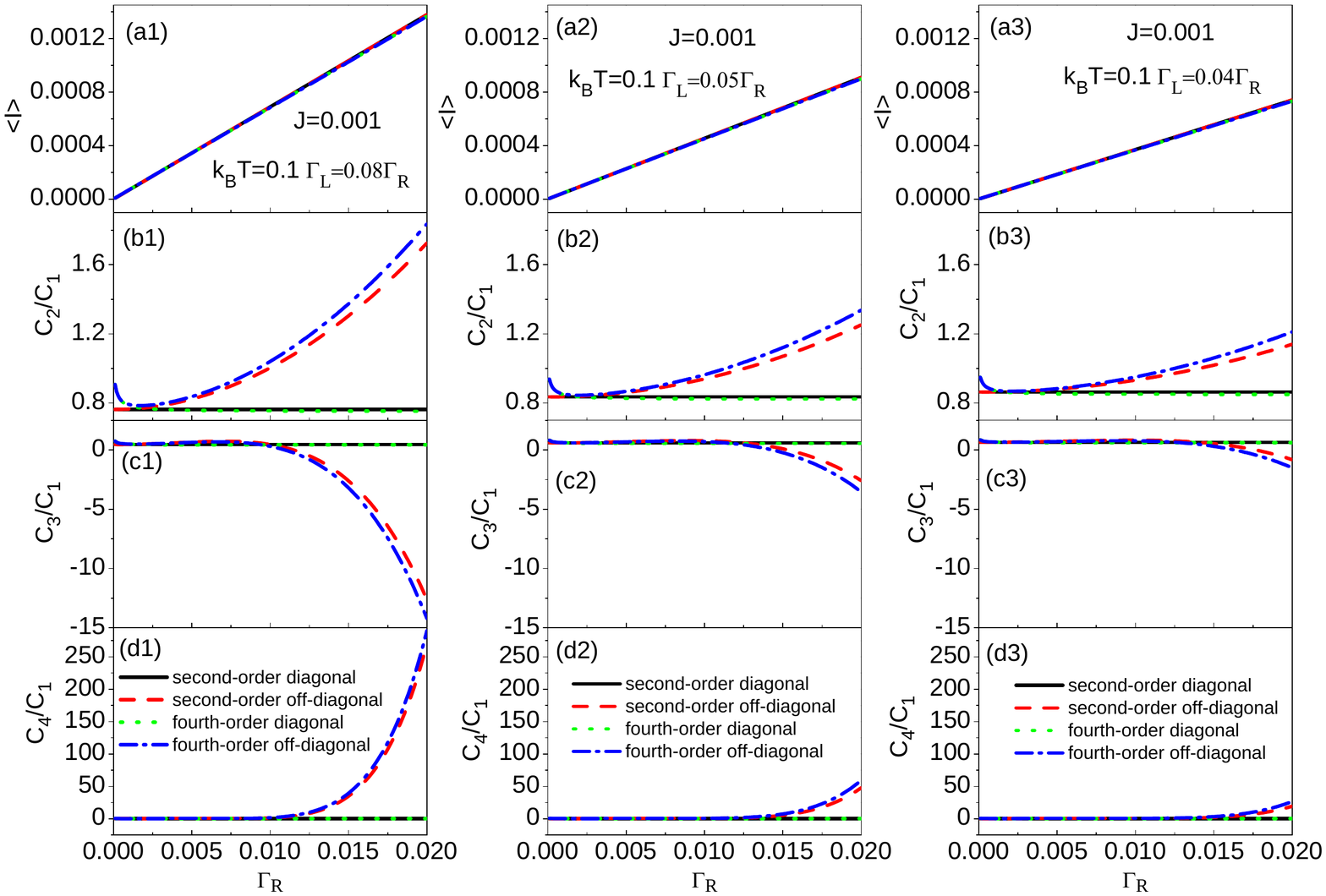}}
\caption{(Color online) The average current $\left\langle I\right\rangle $,
shot noise $C_{2}/C_{1}$, skewness $C_{3}/C_{1}$ and kurtosis $C_{4}/C_{1}$
as a function of the tunneling rate $\Gamma _{R}$ with different values of
the ratio of $\Gamma _{L}$ to $\Gamma _{R}$ at $\Gamma _{L}<\Gamma _{R}$ and
$k_{B}T=0.1$. In the $\Gamma /J\gg 1$ case, the interplay between the
electron cotunneling processes and the quantum coherence determines whether
the super-Poissonian distributions of the shot noise and the kurtosis take
place, and whether the transition of the skewness from positive to negative
values occurs. The notations and the parameters of the QD system are the
same as in Fig. 2.} \label{fig7}
\end{figure*}

\newpage

\begin{figure*}[t]
\centerline{\includegraphics[height=12cm,width=16cm]{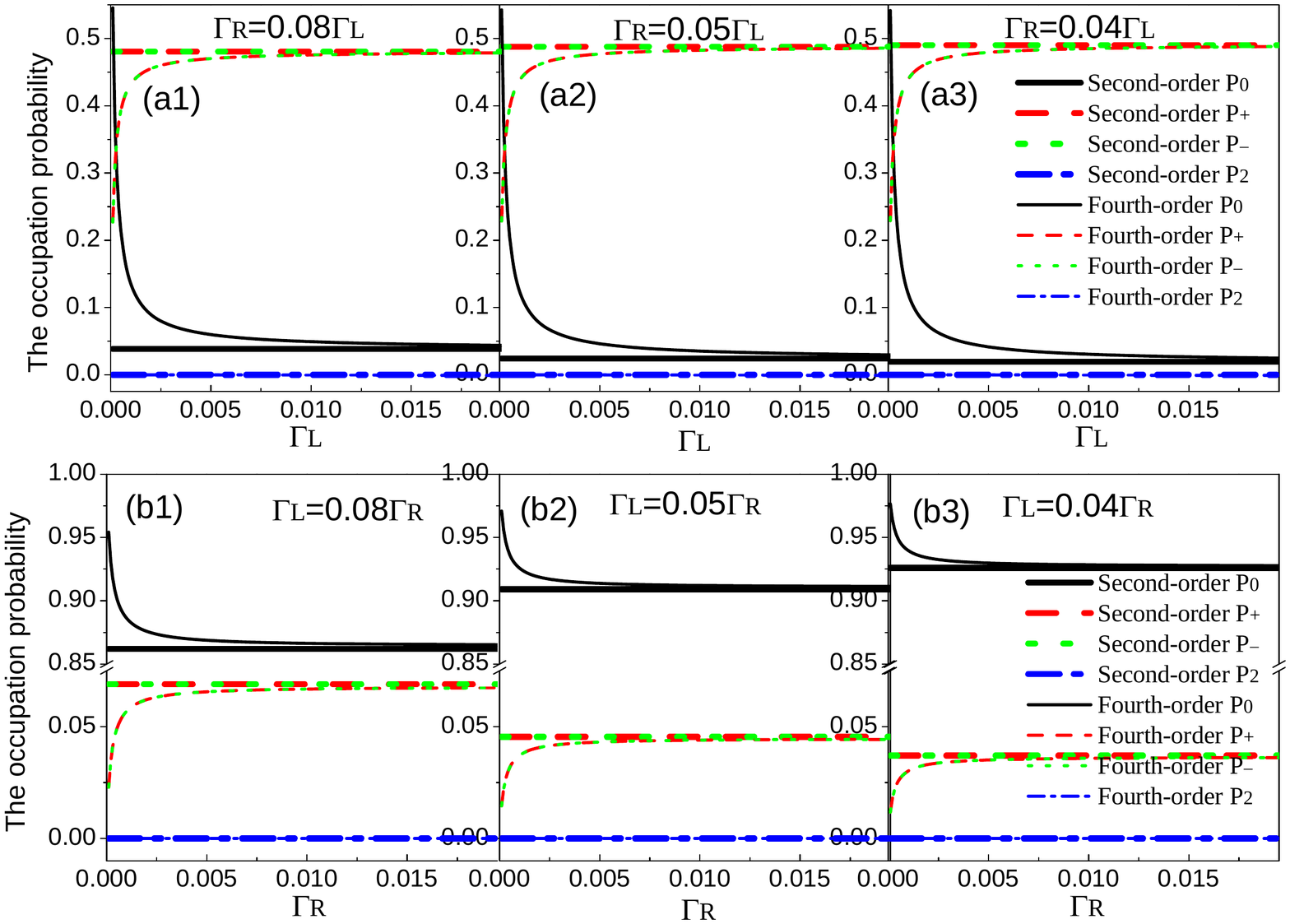}}
\caption{(Color online) The occupation probabilities of the QD's eigenstates
as a function of the tunneling rate $\Gamma _{L}$ ($\Gamma _{R}$) with different values of
the ratio of $\Gamma _{L}$ to $\Gamma _{R}$. The parameters of the QD system
of (a1)-(a3) and (b1)-(b3) are the same as in Figs. 5 and 6, respectively.} \label{fig8}
\end{figure*}

\newpage

\begin{figure*}[t]
\centerline{\includegraphics[height=12cm,width=16cm]{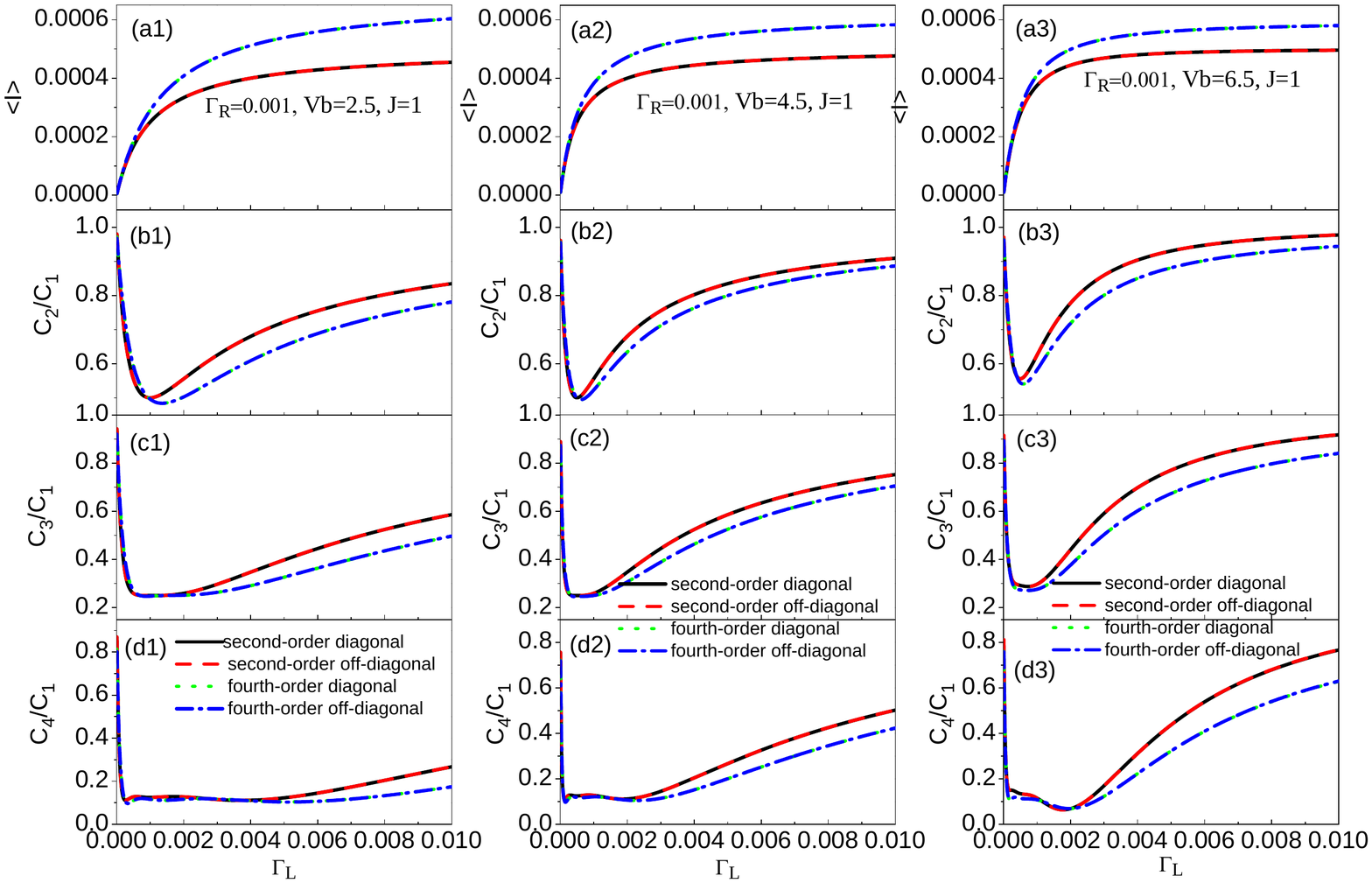}}
\caption{(Color online) The average current $\left\langle I\right\rangle $,
shot noise $C_{2}/C_{1}$, skewness $C_{3}/C_{1}$ and kurtosis $C_{4}/C_{1}$
as a function of the tunneling rate $\Gamma _{L}$ with different bias
voltages $V_{b}$ at $k_{B}T=0.1$ and $\Gamma _{R}=0.001$. Here, the three
fixed bias voltages, under which the different transitions involved in the
electron tunneling, are considered, namely, (1) $V_{b}=2.5$ corresponding to
the transitions between the singly-occupied $\left\vert 1\right\rangle ^{-}$
and empty-occupied eigenstates, (2) $V_{b}=4.5$ corresponding to the
transitions between the singly-occupied $\left\vert 1\right\rangle ^{\pm }$
and empty-occupied eigenstates, (3) $V_{b}=6.5$ corresponding to the
transitions between the singly-occupied $\left\vert 1\right\rangle ^{\pm }$
and empty-occupied eigenstates, and the transitions between the
doubly-occupied $\left\vert 2\right\rangle $ and singly-occupied $\left\vert
1\right\rangle ^{+}$ eigenstates. In the case of $\Gamma _{L}/\Gamma _{R}>1$%
, the electron cotunneling processes have a relatively obvious influence on
the first four order current cumulants. The other notations and the
parameters of the QD system are the same as in Fig. 2.} \label{fig9}
\end{figure*}

\newpage

\begin{figure*}[t]
\centerline{\includegraphics[height=12cm,width=16cm]{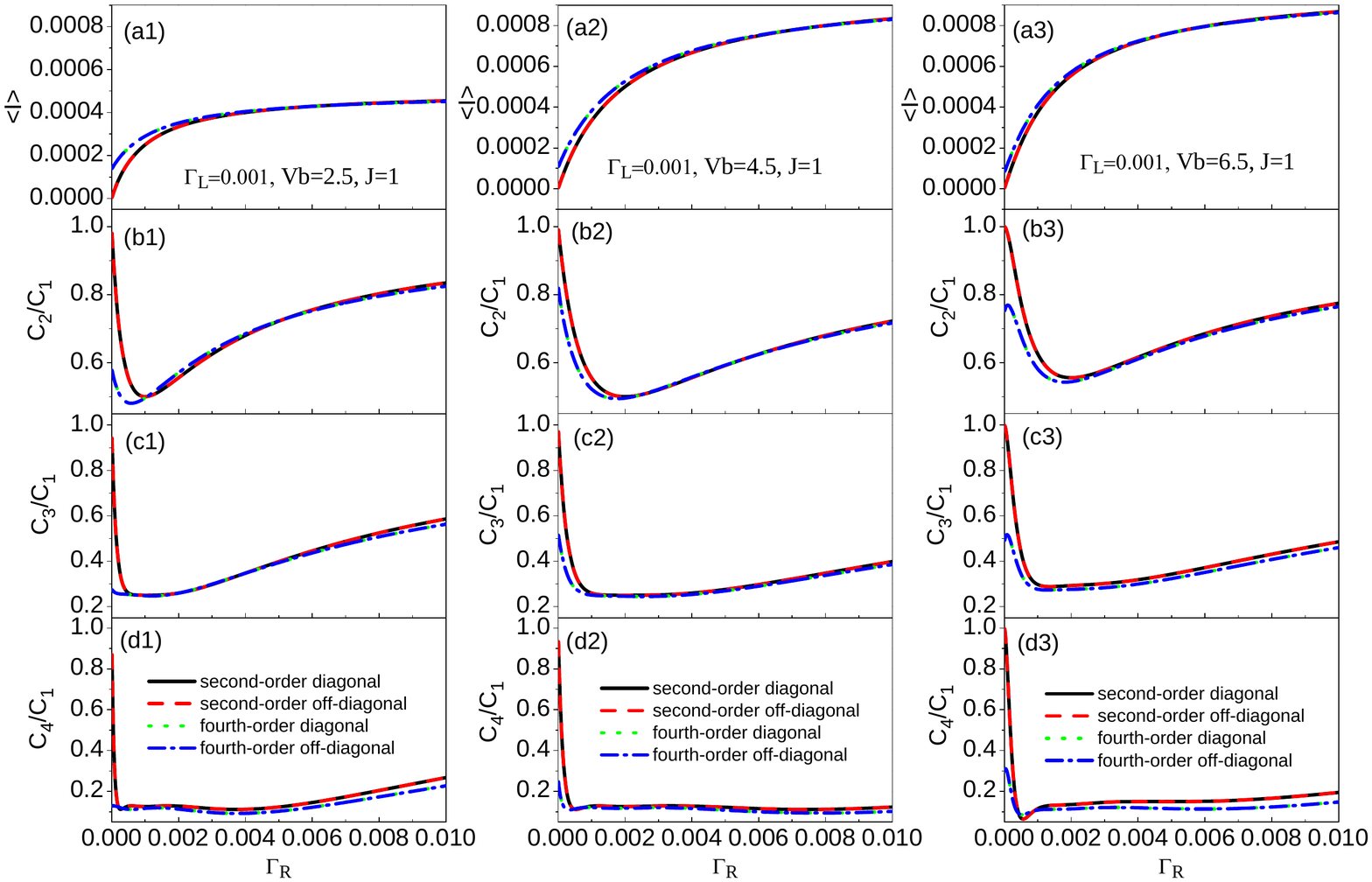}}
\caption{(Color online) The average current $\left\langle I\right\rangle $,
shot noise $C_{2}/C_{1}$, skewness $C_{3}/C_{1}$ and kurtosis $C_{4}/C_{1}$
as a function of the tunneling rate $\Gamma _{R}$ with different bias
voltages $V_{b}$ at $k_{B}T=0.1$ and $\Gamma _{L}=0.001$. In the case of $%
\Gamma _{L}/\Gamma _{R}<1$, the electron cotunneling processes have a slight
influence on the first four order current cumulants. The other notations and
the parameters of the QD system are the same as in Fig. 2.} \label{fig10}
\end{figure*}

\end{document}